\documentclass[letterpaper,twocolumn,10pt]{article}
\usepackage{usenix}

\usepackage{tikz}
\usepackage{amsmath}

\usepackage{filecontents}

\usepackage[utf8]{inputenc} 
\usepackage[T1]{fontenc}    
\usepackage{url}            
\usepackage{hyperref}       
\usepackage{booktabs}       
\usepackage{amsfonts}       
\usepackage{nicefrac}       
\usepackage{microtype}      
\usepackage{xcolor}         

\usepackage{amsmath,amssymb,amsfonts}
\usepackage{mathtools}
\usepackage{cases}
\usepackage{pifont}
\usepackage{multirow}
\usepackage{overpic}
\usepackage{xifthen}
\usepackage{multicol}
\usepackage{cuted}
\usepackage{enumitem}
\usepackage[switch]{lineno}
\usepackage{subcaption}
\usepackage{array}
\newcolumntype{x}[1]{>{\centering\arraybackslash\hspace{0pt}}p{#1}}
\usepackage{makecell}
\usepackage{appendix}
\usepackage{arydshln}
\usepackage[flushleft]{threeparttable}

\newcommand{\grad}{\mathbf{g}}

\newcommand{\tgrad}[1][k]{\widetilde{\Delta}^{(#1)}}

\newcommand{\pfunc}{\phi}

\newcommand{\x}{\mathbf{x}}

\newcommand{\z}{\mathbf{z}}

\newcommand{\X}{\mathbf{X}}
\newcommand{\Y}{\mathbf{Y}}
\newcommand{\Z}{\mathbf{Z}}

\newcommand{\vecb}{\mathbf{b}}

\newcommand{\vecn}{\mathbf{n}}

\newcommand{\stovar}{\xi^{\text{s}}}

\newcommand{\enc}{\mathrm{enc}}
\newcommand{\dec}{\mathrm{dec}}

\newcommand{\V}[1][]{
\ifthenelse{\isempty{#1}}
{\mathbf{V}}
{\mathbf{V}^{(#1)}}}
\newcommand{\vecv}[1][]{
\ifthenelse{\isempty{#1}}
{\mathbf{v}}
{\mathbf{v}^{(#1)}}}

\newcommand{\prob}{\mathbb{P}}

\newcommand{\qw}[1][]{\mathbf{w}}

\newcommand{\epsidx}[1][]{
\ifthenelse{\isempty{#1}}
{\varepsilon^{(t)}}
{\varepsilon^{(t)}_{#1}}}

\newcommand{\rep}{\mathbf{r}}

\newcommand{\w}[1][]{
\ifthenelse{\isempty{#1}}
{\mathbf{w}}
{\mathbf{w}^{(#1)}}}

\newcommand{\argmin}{\mathop{\mathrm{argmin}}\limits} 

\newcommand{\set}{\mathcal{C}}

\newcommand{\ind}[1][]{
\ifthenelse{\isempty{#1}}
{\mathbbm{1}}
{\mathbbm{1} \left(#1\right) }}
\newcommand{\indi}[1][]{
\ifthenelse{\isempty{#1}}
{\mathbbm{1}}
{\mathbbm{1}^{(#1)}}}

\newcommand{\expect}[1][]{
\ifthenelse{\isempty{#1}}
{\mathbb{E}}
{\mathbb{E}\left[#1\right]}}

\newcommand{\RN}[1]{%
  \textup{\uppercase\expandafter{\romannumeral#1}}%
}

\newcommand{\pdist}[1][m]{\mathcal{P}_{m}}

\newcommand{\p}[1][]{
\ifthenelse{\isempty{#1}}
{\mathbf{p}}
{\mathbf{p}^{(#1)}}
}

\newcommand{\sign}{\operatorname{sign}}

\newcommand{\cirone}
{\text{\ding{172}}}
\newcommand{\cirtwo}
{\text{\ding{173}}}
\newcommand{\cirthree}
{\text{\ding{174}}}
\newcommand{\cirfour}
{\text{\ding{175}}}

\newcommand{\intab}[2][0.7]{
\scalebox{#1}{\textrm{#2}}
}

\newtheorem{Example}{Example}
\newtheorem{Definition}{Definition}

\usepackage{tikz}

\newcommand{\highlight}[1]{\vspace{1mm}\noindent{}\textbf{#1}\hspace{1mm}}

\newcommand{\paperfig}[1]{Figure~\ref{#1}}

\newcommand{\zoom}[2][0.7]{  
\scalebox{#1}{#2}
}

\begin{document}

\date{}

\title{\Large \bf Gradient Obfuscation Gives a False Sense of Security in Federated Learning}

\newcommand*{\affaddr}[1]{#1} 
\newcommand*{\affmark}[1][*]{\textsuperscript{#1}}
\newcommand*{\email}[1]{\texttt{#1}}

\author{%
Kai Yue,\affmark[1] Richeng Jin,\affmark[2] Chau-Wai Wong,\affmark[1] Dror Baron,\affmark[1] and Huaiyu Dai\affmark[1]\\
\affaddr{\affmark[1]North Carolina State University; \{kyue, chauwai.wong, dzbaron, hdai\}@ncsu.edu } \\
\affaddr{\affmark[2]Zhejiang University; richengjin@zju.edu.cn}\\
}


\maketitle
\begin{abstract}
Federated learning has been proposed as a privacy-preserving machine learning framework that enables multiple clients to collaborate without sharing raw data. 
However, client privacy protection is not guaranteed by design in this framework.  
Prior work has shown that the gradient sharing strategies in federated learning can be vulnerable to data reconstruction attacks.
In practice, though, clients may not transmit raw gradients considering the high communication cost or due to privacy enhancement requirements. 
Empirical studies have demonstrated that gradient obfuscation, including intentional obfuscation via gradient noise injection and unintentional obfuscation via gradient compression, can provide more privacy protection against reconstruction attacks. 
In this work, we present a new reconstruction attack framework targeting the image classification task in federated learning.
We show how commonly adopted gradient postprocessing procedures, such as gradient quantization, gradient sparsification, and gradient perturbation may give a false sense of security in federated learning. 
Contrary to prior studies, we argue that privacy enhancement should not be treated as a byproduct of gradient compression.  
Additionally, we design a new method under the proposed framework to reconstruct images at the semantic level. 
We quantify the semantic privacy leakage and compare it with conventional image similarity scores. 
Our comparisons challenge the image data leakage evaluation schemes in the literature. 
The results emphasize the importance of revisiting and redesigning the privacy protection mechanisms for client data in existing federated learning algorithms. 
\end{abstract}

\section{Introduction}
The past decade has seen a growing demand for a large amount of training data for deep learning, as well as the increased storage and computational capabilities of edge devices.    
Federated learning has emerged as a privacy-preserving framework under which multiple participants jointly solve a machine learning problem~\cite{kairouz2021advances,wang2021a}.
In a classical federated network, a central server broadcasts a global model to selected clients and collects model updates without directly accessing raw data. 
Generic solutions such as federated averaging (FedAvg)~\cite{mcmahan2017communication} have been proposed with different variants.
In FedAvg, a server transmits a global model to participants based on a predefined client sampling strategy. 
On the client side, the model will be locally optimized with decentralized private training data. 
Once the models are transmitted back to the server, an updated global model is constructed by averaging all received local models. 
During the whole process, raw data will not be exchanged. 
To reduce the communication cost and improve the model accuracy, especially when clients' data are not independent and identically distributed (IID), gradient quantization and personalized client cost functions have been incorporated into FedAvg~\cite{reisizadeh2020fedpaq, li2020fedprox}. 

Even without direct access to the raw data, client privacy is not guaranteed in the aforementioned FedAvg framework. 
Recent studies have shown that client data can be reconstructed based on gradient inversion attack~\cite{zhu2019deep, zhao2020idlg}. 
Specifically, Zhu et al.~\cite{zhu2019deep} proposed an attack method that can reconstruct pixel-wise accurate images and token-wise matched texts. 
This method was further improved to recover high-resolution images with increased batch sizes~\cite{geiping2020inverting, yin2021see}. 
Meanwhile, various attack evaluation frameworks have been proposed to quantify the privacy loss based on image similarity scores calculated between the reconstructed images and clients' raw images~\cite{wei2020framework, huang2021evaluating}. 

To achieve provable differential privacy, researchers have studied adding noise to gradients without significantly reducing the model utility~\cite{wei2021user, yu2021do}. 
Nevertheless, it is not clear how differentially private gradients can defend against well-designed data reconstruction attacks~\cite{zhang2020secret}.
In addition to gradient perturbation, empirical studies demonstrated that gradient compression can serve as a good deterrence to the reconstruction attack~\cite{zhu2019deep, wei2020framework}. 
Recent studies suggest that low-bit gradient quantization can be used as effective defenses~\cite{zhang2022survey}.
Since gradient compression is proposed to reduce the communication cost, privacy protection is not considered by design. 
Quantifying the privacy gain of various gradient obfuscation procedures, including intentional ones such as differential privacy through noise injection and unintentional ones such as gradient compression, remains an open problem. 

In this study, we show that the aforementioned gradient obfuscation strategies may give a false sense of security. 
We investigate the image classification task with high-resolution input and improve the image reconstruction attack.
We summarize our contributions as follows:
\begin{itemize}
    \item We propose a reconstruction attack framework with improved reconstruction quality in federated learning.
          Contrary to previous studies, we show that gradient compression, such as quantized stochastic gradient descent (QSGD)~\cite{alistarh2017qsgd} and Top-$k$ sparsification~\cite{aji2017sparse}, may not be treated as effective defenses to prevent privacy leakage. 
    \item We evaluate several existing defense schemes that perturb the model updates or features.
          The attack results show that these defenses can still be vulnerable to adversaries, thus advocating the importance of more effective privacy protection designs in federated learning. 
    \item We explore an image reconstruction attack at a semantic level and measure the corresponding image privacy leakage. 
          Although the reconstructions do not necessarily match the original images pixel by pixel, the private information can still be exploited by adversaries.   
          Our work is a step toward a more thorough understanding of the privacy leakage problem in federated learning.
\end{itemize}

The rest of this paper is organized as follows. 
Section~\ref{section:related_work} reviews the relevant work. 
Section~\ref{section:preliminaries} describes the background of federated learning, including the concepts, gradient postprocessing tools, and data reconstruction attacks. 
We present the reconstruction framework in Section~\ref{section:rog} and evaluate three existing defense schemes in Section~\ref{section:current_defense}.
Section~\ref{section:semnatic_attack} introduces the attack at the semantic level and the corresponding evaluation metric. 
Finally, we conclude the paper and discuss the future work in Section~\ref{section:discussion}.  

\section{Related Work}\label{section:related_work}

\subsection{Federated Learning and Client Privacy}
In federated learning, a server coordinates the learning procedure and generally has access to the model structure and parameters.
In this work, the investigated threat model is an \emph{honest-but-curious} server that follows the FedAvg framework to aggregate client updates, while inspecting the private information without interfering with the training.
The attacks concerning data privacy and model confidentiality in federated learning can be categorized into three types: membership inference attack, model inversion/data reconstruction attack, and property inference attack~\cite{jere2020taxonomy, rigaki2020survey}. 

In spite of its popularity, FedAvg and its variants have exhibited vulnerabilities in protecting client data privacy~\cite{lyu2020threats}. 
Recent studies have shown a particular interest in model inversion attacks. 
Zhu et al.~\cite{zhu2019deep} proposed the \textit{deep leakage from gradient} (DLG) algorithm, in which a dummy input batch of data is updated iteratively to match the shared gradients. 
They demonstrated that DLG could reconstruct pixel-wise accurate images and token-wise matched texts. 
This attack has been further developed by revealing labels via analytical methods~\cite{zhao2020idlg, wainakh2021label, yin2021see}. 
Researchers have proposed to reconstruct high-resolution inputs by employing the image prior and improving the design of the cost functions~\cite{jeon2021gradient, yin2021see, geiping2020inverting}. 
For example, Jeon et al.~\cite{jeon2021gradient} updated the parameters of a generative model to obtain reconstructed results. 
Meanwhile, it has been shown that the modern Transformer model adopted in language tasks can also be compromised under the data reconstruction attack~\cite{deng2021tag}. 
From a mathematical viewpoint, researchers have demonstrated that DLG is equivalent to solving a system of equations~\cite{pan2020theory, qian2020minimal}. 
The data reconstruction attack can also be interpreted from a Bayesian perspective~\cite{balunovic2021bayesian}.  

Our work is closely related to those studies focused on gradient inversion attacks in federated learning~\cite{zhu2019deep,zhao2020idlg,geiping2020inverting,yin2021see}. 
We propose a new attack framework toward high-resolution image data reconstruction in federated learning. 
By assuming a successful analytical label recovery and improving the design to reduce the redundancy in the unknowns, we increase the quality of reconstructed images and the efficiency of the attack algorithm.  
Compared to prior work that does not consider client local update~\cite{yin2021see} or does not consider a batch of input data for high-resolution images~\cite{zhu2019deep,geiping2020inverting}, we show that an adversary can still reconstruct client data in a more realistic federated learning setting.

\subsection{Attack Evaluation and Privacy Leakage}  
In the literature, various evaluation metrics are adopted as proxies to measure privacy leakage. 
Pioneering work~\cite{zhu2019deep,geiping2020inverting, wei2020framework} used a conventional image similarity score or the distance between original and reconstructed images as the evaluation metric, including the mean squared error (MSE), peak signal-to-noise ratio (PSNR)~\cite{gonzalez2014digital}, and structural similarity index measure (SSIM)~\cite{wang2004image}. 
In addition, Wei et al.~\cite{wei2020framework} proposed the attack success rate, namely, the percentage of successfully reconstructed training data. 
Follow-up studies~\cite{yin2021see, huang2021evaluating} further leveraged the neural network based image perceptual scores, such as learned perceptual image patch similarity (LPIPS)~\cite{zhang2018unreasonable} as the attack evaluation metric. 
LPIPS is known to emulate humans' perception well so it has been increasingly popular for evaluating the quality of reconstructed images. 
Compared to other indicators, such as differential privacy budget of the algorithm or mutual information between the training examples and gradient~\cite{mo2021quantifying, liu2021quantitative}, LPIPS is more intuitive and interpretable. 
We will use LPIPS as one of the evaluation metrics.

In this work, we demonstrate how an attacker can reveal the high-level semantics in the raw image, which may not be captured by the image similarity scores such as SSIM or PSNR.
We find that there exists ambiguity in telling whether an attack is successful or not. 
These results raise concerns about the existing evaluations and highlight the importance to revisit the privacy leakage issue in federated learning. 
We further propose a method to evaluate semantic privacy loss.

\subsection{Privacy Enhancement and Defense}
Preserving privacy is a priority in designing federated learning algorithms. 
Homomorphic encryption provides privacy protection by aggregating the ciphertexts directly~\cite{gilad2016cryptonets}. 
However, the additional computational cost and communication overhead could dominate the training procedure~\cite{zhang2020batchcrypt}.  
Meanwhile, differentially private machine learning has been developed to achieve provable privacy guarantees~\cite{abadi2016deep}. 
After clipping the gradient and adding Gaussian noise, user-level membership privacy can be enhanced at the cost of the model accuracy~\cite{wei2021user, wei2021gradient}. 
More sophisticated schemes add noise to the manifold representation of gradients~\cite{kerkouche2021compression, yu2021do}, where the model accuracy can be greatly improved given the same privacy protection with reduced noise power.  
However, the vulnerabilities can still be exploited by model inversion attacks as differential privacy does not imply protection against reconstruction attacks or Bayesian restoration targeting attribute privacy~\cite{gu2021federated, zhang2020secret}.
Similar issues also exist in those defense methods that put attention on a particular type of attack~\cite{boutet2021mixnn}.  
Other works focused on the protection provided by a trusted shuffling server~\cite{girgis2021shuffled,cheng2021separation} or secure aggregation protocols~\cite{bonawitz2016practical}. 
These methods preserve privacy while increasing the system complexity drastically. 

In this study, we perform attacks on various gradient obfuscation defense schemes, including gradient compression~\cite{wei2020framework}, differentially private training~\cite{wei2021gradient}, and representation perturbation~\cite{yu2021do}.
We do not consider cryptography based defense~\cite{zhang2020batchcrypt} and other more advanced solutions, including blockchain-based decentralized optimization or federated network topology modification~\cite{nguyen2021federated}.
\section{Preliminaries}\label{section:preliminaries}
Consider a federated learning architecture optimized with FedAvg, which is a backbone of commonly-adopted federated learning algorithms.  
Denote the $m$th client's dataset by $\mathcal{D}_m =  \{(\x_{m,i},y_{m,i})\}_{i=1}^{N_m}$, where the $i$th example $(\x_{m,i},y_{m,i})$ contains an input-output pair drawn from a distribution $\mathcal{P}_m$.
The local objective function $f_m$ is defined as the empirical risk over the local data:
\begin{equation}
    f_m(\w) \triangleq \frac{1}{N_m} \sum_{i=1}^{N_m} \ell(\w; \x_{m,i}, y_{m,i}),
\end{equation}
where $\ell$ is a sample-wise loss function quantifying the error of the model with a weight vector $\w \in \mathbb{R}^{d}$ estimating the label $y_{m,i}$ given an input $\x_{m,i}$. 
Federated learning optimizes the following problem:
\begin{equation}
    \min_{\w \in \mathbb{R}^{d}} f(\w) = \frac{1}{M} \sum_{m=1}^{M} f_{m}(\w),
\end{equation}
where $M$ is the total number of clients.
In FedAvg, the server will select a subset $\set_k$ of clients and broadcast a global model $\w[k]$ in each communication round $k$.
Once the model is received, the $m$th client will initialize a local model $\w[k,0]_m$ and optimize it with multiple gradient descent steps. 
For the gradient descent based method, the local model updated at step $t$ may be formulated as 
\begin{equation}
    \w[k,t]_{m}\! = \w[k,0]_m \!-\! \frac{\eta}{N_m} \sum_{u=0}^{t-1}\!\sum_{i=1}^{N_m}\! \nabla \ell (\w[k,u]_m; \x_{m,i}, y_{m,i}), 
\end{equation}
where $\eta$ is the learning rate. 
For simplicity, in this work we also call the weight difference $\w[k,0]_m - \w[k,t]_m$ the gradient. 
After $\tau$ local update steps, the client will choose to transmit an obfuscated gradient $\tgrad[k]$ based on a postprocessing function $\pfunc(\cdot)$, i.e., 
\begin{equation}
    \tgrad[k]_{m} = \pfunc(\w[k,0]_{m} - \w[k,\tau]_{m}). 
\end{equation}

\highlight{Threat model. }
We consider an honest-but-curious server as an attacker. 
The goal of the attacker is to inspect sensitive information of client private data without accessing them directly. 
Specifically, the server may reconstruct training examples that are close to raw ones. 
The attacker knows the shared model $\w[k]$ and transmitted gradients $\tgrad[k]_{m}$'s, while it does not modify the model or gradients.  
It can also access public data and pretrained neural network models such as an image-denoising network. 
In general, the attacker has sufficient computational resources and memory. 
We do not consider a stronger attacker that can modify the gradients or model weights.
It has been shown that when the integrity of the model/gradient is compromised, the data can be trivially reconstructed~\cite{fowl2021robbing, wen2022fishing}.

We characterize the postprocessing function $\pfunc(\cdot)$ as follows. 
First, we describe several examples of $\pfunc(\cdot)$ that are designed to improve communication efficiency.
\begin{Example}
(SignSGD Compression). 
To save communication cost, the client will use a compression function to reduce the size of gradients. 
For example, one can use the low precision quantization such as signSGD \cite{bernstein2018signsgd}, i.e., 
\begin{equation}
    \pfunc_{\text{sign}}(\grad) = \sign(\grad).
\end{equation}    
\end{Example}
Other commonly adopted compression schemes include FedPAQ \cite{reisizadeh2020fedpaq} and Top-$k$ sparsification \cite{aji2017sparse}. 
We give the example of gradient quantization and sparsification as follows.
We use $\pfunc_{\text{q}}$ and $\pfunc_{\text{qsgd}}$ to represent a deterministic uniform quantizer and a stochastic QSGD~\cite{alistarh2017qsgd} quantizer, respectively. 
\begin{Example}
(Gradient Quantization). 
A client can use a quantizer to compress the gradients. 
Given an input $\grad$, a quantization operator $Q$ will process each entry $g_i$ as follows:
\begin{equation}\label{eq:quant_operator}
    Q\left(g_{i}\right)=\frac{\kappa}{s}\|\grad\|_{p} \cdot \operatorname{sign}\left(g_{i}\right) \cdot \xi_{i}(\grad,\, s),
\end{equation}
where $\kappa$ is a scaling factor, $s$ is a predefined parameter and the number of representation levels is equal to $2s+1$, $\|\grad\|_p$ is the $\ell_{p}$ norm of the input vector $\grad$, and $\xi_i(\grad, s)$ is an integer value representing the unsigned quantized level of the $i$th coordinate of the input vector $\grad$.
For a stochastic quantizer $\pfunc_{\text{qsgd}}$, we set $\xi_i$ in \eqref{eq:quant_operator} by $\stovar_i$ defined as follows:
\begin{equation}\label{def:normalization_qsgd}
    \stovar_i(\grad, s) = \left\{
        \begin{array}{l @{\; \;} l}
        l &  \text { with prob. } 1-p_{i}, \\[5pt]
        l + 1 & \text { with prob. } p_{i} = \frac{s|e_i|}{\kappa \|\grad\|_p} - l,
    \end{array}\right.
\end{equation}
where $l \in [0,s)$ is an integer and $\frac{|e_{i}|}{\kappa \|\grad\|_{p}} \in[l / s,(l+1) / s]$. 
\end{Example}
\begin{Example}
(Gradient Sparsification).
In Top-$k$ compression~\cite{aji2017sparse}, the client selects the $k$ largest components of the gradient in absolute values and zeros out other entires.   
We denote this function by $\pfunc_{\text{topk}}$. 
\end{Example}
Next, we state the definition of differential privacy and give the example of noisy gradients.  
\begin{Definition}
(Differential Privacy).
A randomized mechanism $\mathcal{M}: \mathcal{D} \rightarrow \mathcal{R}$ with domain $\mathcal{D}$ and range $\mathcal{R}$ satisfies $(\varepsilon, \delta)$-differential privacy if, for any two adjacent inputs $d, d^{\prime} \in \mathcal{D}$ and for any subset of outputs $S \subseteq \mathcal{R}$ it holds that
\begin{equation}
    \prob[\mathcal{M}(d) \in S] \leq e^{\varepsilon} \prob\left[\mathcal{M}\left(d^{\prime}\right) \in S\right]+\delta. 
\end{equation}
\end{Definition}

\begin{Example} 
(Noisy Gradient). 
To achieve a provable differential privacy guarantee, the client will inject additive Gaussian noise to the gradient~\cite{abadi2016deep, wei2021user}. 
Suppose the input $\grad$ is bounded,  the postprocessing function may be written as 
\begin{equation}
    \pfunc_{\text{dp}}(\grad) = \grad + \vecn,
\end{equation}
where each entry of $\vecn$ is an independent Gaussian variable. 
The privacy protection increases with the noise power~\cite{wei2021gradient}. 
\end{Example}

We now review the data reconstruction attack methods of DLG~\cite{zhu2019deep} and \textit{inverting gradient} (InvertGrad)~\cite{geiping2020inverting}.
Consider a client $m$ holding image data $\{\X_{m,i}\}_{i=1}^{N_m}$, where each matrix $\X_{m,i} \in \mathbb{R}^{h \times w}$ denotes an image with height $h$ and width $w$. 
Here, we assume grayscale images for the simplicity of presentation unless otherwise specified.
An adversarial server is interested in reconstructing $\{\hat{\X}_{m,i}\}_{i=1}^{N_m}$ that is close to the client raw data \footnote{We note that label information $y_{m,i}$ is also sensitive in supervised learning. 
Existing studies~\cite{yin2021see, wainakh2021label} have designed efficient algorithms to unveil labels with high accuracy. 
In this work, we assume the label information is readily derived via analytical approaches.
This assumption is also adopted in prior attack work~\cite{geiping2020inverting}. }.   
In DLG \cite{zhu2019deep}, the attacker starts from some input $\X'_{m,i}$, calculates the dummy gradient $\Delta'^{(k)}_m$, and solves the following optimization problem to match with the shared gradient $\tgrad[k]_{m}$: 
\begin{equation}\label{eq:optim_problem}
    \{\hat{\X}_{m,i}\}_{i=1}^{N_m} = \argmin_{\{\X'_{m,i}\}_{i=1}^{N_m}} \| \Delta'^{(k)}_m - \tgrad[k]_{m} \|^{2}. 
\end{equation}
In InvertGrad~\cite{geiping2020inverting}, the cost function is replaced by cosine distance and the attack optimization is formulated as follows:
\begin{equation}
    \{\hat{\X}_{m,i}\}_{i=1}^{N_m} = \argmin_{\{\X'_{m,i}\}_{i=1}^{N_m}} - \frac{\langle \Delta'^{(k)}_m, \tgrad[k]_{m} \rangle}{\|\Delta'^{(k)}_m\|\| \tgrad[k]_{m}\|} + \frac{1}{N_m} \sum_{i=1}^{N_m} \mathcal{L}_{\text{TV}}(\X'_{m,i}),
\end{equation}
where $\mathcal{L}_{\text{TV}}$ is the total variation given by 
\begin{equation}
    \mathcal{L}_{\text{TV}}(\X)=\sum_{i, j} \sqrt{\left|X_{i+1, j}-X_{i, j}\right|^{2}+\left|X_{i, j+1}-X_{i, j}\right|^{2}}, 
\end{equation}
where $X_{i,j}$ denotes the $(i,j)$th entry of the matrix $\X$. 

In a practical federated learning setting, when the gradients are averaged over multiple training examples and local iterations, the attack optimization problem is still difficult to solve \cite{geiping2020inverting, yin2021see}.
The effectiveness of these attack methods becomes even more questionable when the gradients are postprocessed by the function $\pfunc(\cdot)$, leaving vague conclusions on the privacy protection provided by federated learning.

\section{Reconstruction From Obfuscated Gradient}\label{section:rog}

\begin{figure}[!tb]
    \centering
    \begin{overpic}[width=0.48\textwidth]{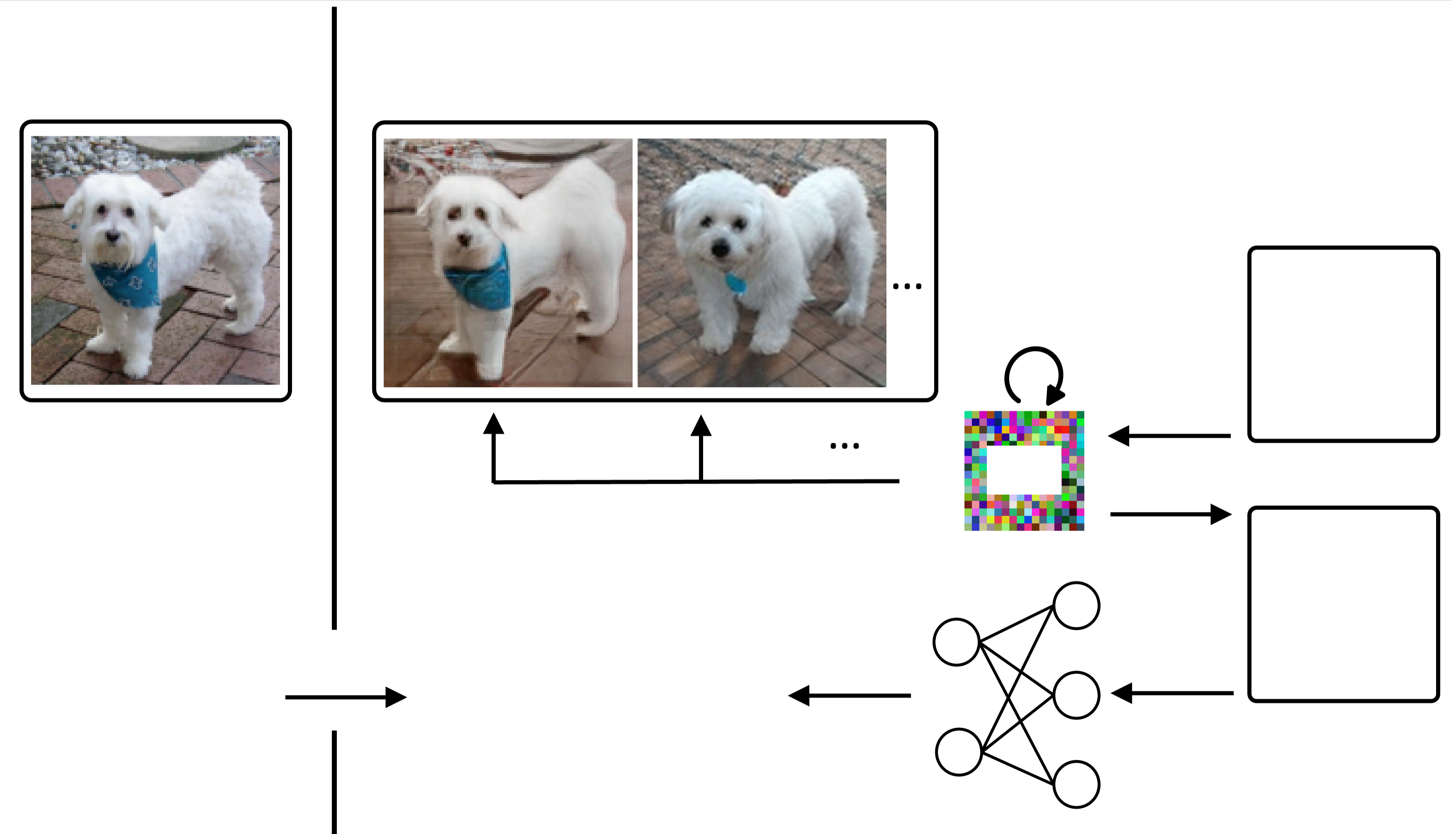}
    \put(0.5, 50){\intab{private data $\X_{m,i}$}}
    \put(0, 20){\intab[0.65]{$\centerdot$ local optimization}}
    \put(0, 15){\intab[0.65]{$\centerdot$ compute $\Delta_m^{(k)}$}}
    \put(0, 10){\intab[0.65]{$\centerdot$ $\tgrad[k]_{m} = \pfunc(\Delta_m^{(k)})$}}
    \put(19, 11){\intab[0.65]{upload}}

    \put(30, 50){\intab{reconstructed data $\hat{\X}_{m,i}$'s}}
    \put(2, 55){\intab[0.84]{\textbf{client side}}}
    \put(26, 55){\intab[0.84]{\textbf{server side}}}
    \put(74.5, 29){\intab{\cirone}}
    \put(89, 33){\intab{$\X'_{m,i}$}}

    \put(77.5, 29){\intab[0.65]{$\enc(\cdot)$}}
    \put(74.5, 23.2){\intab{\cirtwo}}
    \put(77.5, 23.2){\intab[0.65]{$\dec(\cdot)$}}

    \put(89, 15){\intab{$\hat{\X}'_{m,i}$}}
    \put(76.5, 11){\intab{input}}
    \put(53, 11){\intab{gradient}}
    \put(31, 9){\intab{$\displaystyle \min_{\Z_{m,i}} \| \Delta'^{(k)}_m - \tgrad[k]_{m} \|^{2}$}}

    \put(28, 9){\intab{\cirthree}}
    \put(69, 35){\intab{\cirthree}}
    \put(72, 35){\intab{update}}
    \put(31, 21){\intab{\cirfour}}
    \put(34.5, 21){\intab{decode \& postprocess}}

    \put(67, 24.3){\intab{$\Z_{m,i}$ }}
\end{overpic}
\caption{Schematic of the proposed attack method. 
The attacker first encodes an initial image $\X'_{m,i}$ into a low-dimensional representation $\Z_{m,i}$ to reduce the number of unknowns. 
Second, the attacker calculates the gradient by decoding the representation $\Z_{m,i}$ to image $\hat{\X}'_{m,i}$ and feeding it to the target neural network model. 
Third, the representation $\Z_{m,i}$ is updated by solving an optimization problem to minimize the discrepancy between the dummy gradient $\Delta'^{(k)}_m$ and client gradient $\tgrad[k]_m$. 
The final reconstruction results are obtained via decoding and postprocessing. 
Reconstructions can be different based on different postprocessing tools or optimization procedures.  
\label{fig:rog}
}
\end{figure}
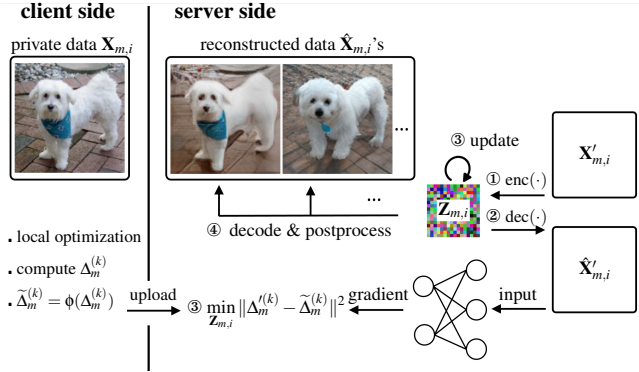

In this section, we present our framework to \textbf{\underline{r}}econstruct image data from the \textbf{\underline{o}}bfuscated \textbf{\underline{g}}radient (ROG).
The schematic is illustrated in Figure~\ref{fig:rog}, and the details are given in the following sections. 
For the simplicity of presentation, we first discuss the reconstruction attack on a single image $\X_{m,i}$.
The method can be applied to the reconstruction of a batch of data. 

\subsection{Attack Scheme}
The proposed attack scheme uses four steps to reconstruct client image data. 
We will first introduce each step in its general form, and discuss the realizations in Section~\ref{section:realization}.

\textbf{First}, the attacker derives the label $y_{m,i}$ and selects an initial image $\X'_{m,i}$.
The image can be randomly initialized, for example, by using independent random variables drawn from a Gaussian distribution or a uniform distribution. 
The initialization can also be implemented as repetitive patterns or a natural image~\cite{wei2020framework}. 
After the initialization, instead of looking for the numerical solutions of \eqref{eq:optim_problem} directly, we introduce an encoding step to preprocess the image first.    
From the perspective of solving the system of equations, the solution will be easier to determine if the number of unknowns decreases.
To reduce the correlations among different coordinates of input variables, we project the image $\X'_{m,i}$ to a low-dimensional representation $\Z_{m,i}$, which is formulated as 
\begin{equation}
    \Z_{m,i} = \enc(\X'_{m,i}).
\end{equation}
The encoding function $\enc(\cdot)$ can be implemented intuitively as a lossy bicubic downsampling function~\cite{gonzalez2014digital}, or as complicated as a neural network encoder~\cite{baldi2012autoencoders}. 
In the \textbf{second} step, the attacker applies the decoding function $\dec(\cdot)$ to map the representation $\Z_{m,i}$ back to an image $\hat{\X}'_{m,i}$, which is then fed into the federated learning model together with the label $y_{m,i}$ to compute the gradient $\Delta'^{(k)}_m$. 
\textbf{Third}, the attacker solves the following minimization problem (e.g., using Adam~\cite{kingma2015adam}) and update the compact representation $\Z_{m,i}$ accordingly:
\begin{equation}\label{eq:new_optim_problem}
    \Z^\star_{m,i} = \argmin_{\Z_{m,i}} \| \Delta'^{(k)}_m - \tgrad[k]_{m} \|^2. 
\end{equation}
\textbf{Fourth}, after the optimization terminates when the error is below a threshold or after some predefined number of iterations, we decode $\Z^\star_{m,i}$ and use postprocessing tools to enhance the image quality. 
Next, we will present the realizations under the attack framework and make comparisons with existing work. 

\subsection{Realizations and Comparisons}\label{section:realization}
\begin{figure}[!tb]
    \begin{overpic}[width=\linewidth]{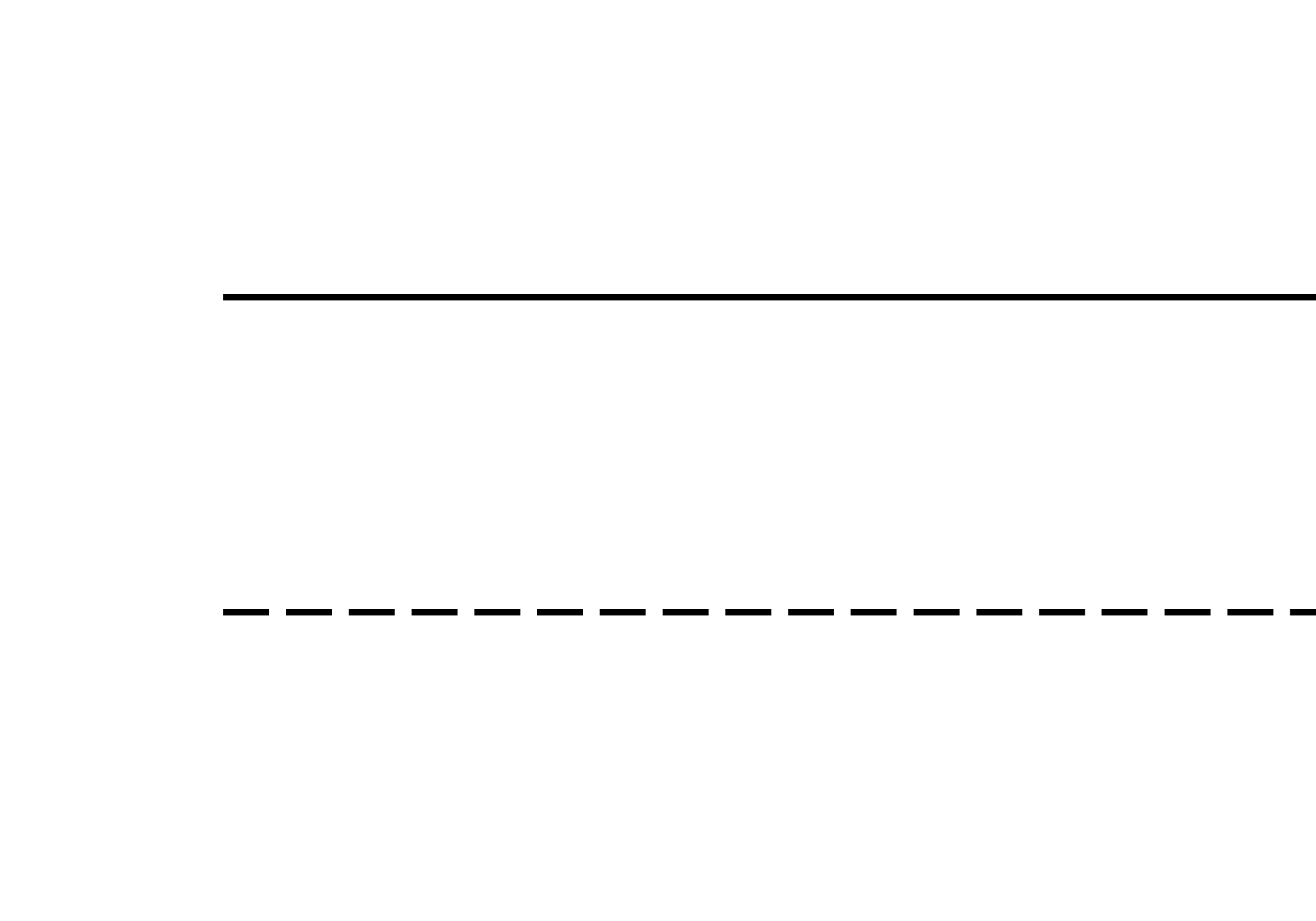}
    \put(17, 49){\includegraphics[width=0.2\linewidth]{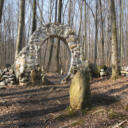}}
    \put(38, 49){\includegraphics[width=0.2\linewidth]{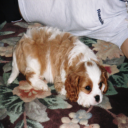}}
    \put(59, 49){\includegraphics[width=0.2\linewidth]{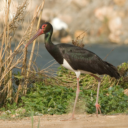}}
    \put(80., 49){\includegraphics[width=0.2\linewidth]{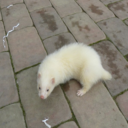}}

    \put(17, 25.6){\includegraphics[width=0.2\linewidth]{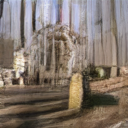}}
    \put(38, 25.6){\includegraphics[width=0.2\linewidth]{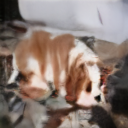}}
    \put(59, 25.6){\includegraphics[width=0.2\linewidth]{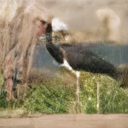}}
    \put(80, 25.6){\includegraphics[width=0.2\linewidth]{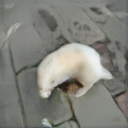}}

    \put(17, 1.){\includegraphics[width=0.2\linewidth]{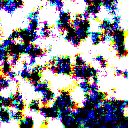}}
    \put(38, 1.){\includegraphics[width=0.2\linewidth]{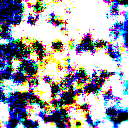}}
    \put(59, 1.){\includegraphics[width=0.2\linewidth]{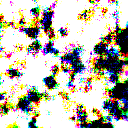}}
    \put(80, 1.){\includegraphics[width=0.2\linewidth]{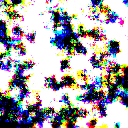}}

    \put(0, 58){\zoom{Raw Images}}
    \put(0, 36){\zoom{ROG Attack}}
    \put(0, 32){\zoom{LPIPS $0.210$}}
    \put(0, 13){\zoom{InvertGrad}}
    \put(0, 9){\zoom{LPIPS $0.717$}}
    \end{overpic}
    \vspace*{-15pt}
    \caption{Reconstructed images using two attack schemes on LeNet under FedAvg. 
    ROG reconstruction $\hat{\X}_{m,i}$'s are visually similar to the raw images $\X_{m,i}$'s. 
    In comparison, the attack algorithm InvertGrad fails to recover meaningful private visual information. 
    \label{fig:attack_fedavg}}
\end{figure}
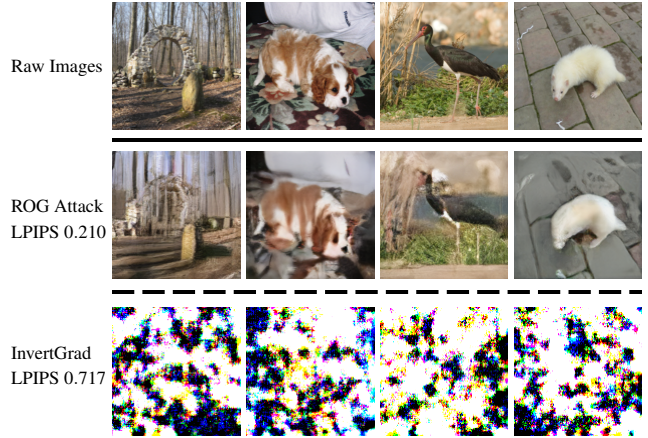

One implementation of the ROG framework is given as follows. 
We use independent random variables from uniform distribution $U(0,1)$ to initialize the reconstruction image $\X'_{m,i}$. 
To ensure the computational efficiency and avoid vanishing gradient in the backpropagation, we choose bicubic downsampling as the encoding function $\enc(\cdot)$, with a scaling factor of $4$. 
The decoding function $\dec(\cdot)$ corresponds to the bicubic upsampling with the same scaling factor.
Finally, we treat the postprocessing in the last step as an image enhancement task. 
In particular, we randomly apply one of the downsampling approaches, including nearest neighbor, bilinear scaling, and bicubic scaling,  and add Gaussian noise to the training examples in the ImageNet dataset~\cite{deng2009imagenet}.
This procedure approximates the effect of the image degradation at different levels~\cite{zhang2021designing}. 
Later, a neural network is optimized with the synthetic dataset to enhance the image quality. 
More details of the experimental setups and implementations can be found in Appendix~\ref{section:implementation}. 

\begin{figure}[!tb]
    \subcaptionbox{\label{subfig:ori_fullbatch}}[0.495\linewidth]{
    \begin{overpic}[width=\linewidth, height=\linewidth]{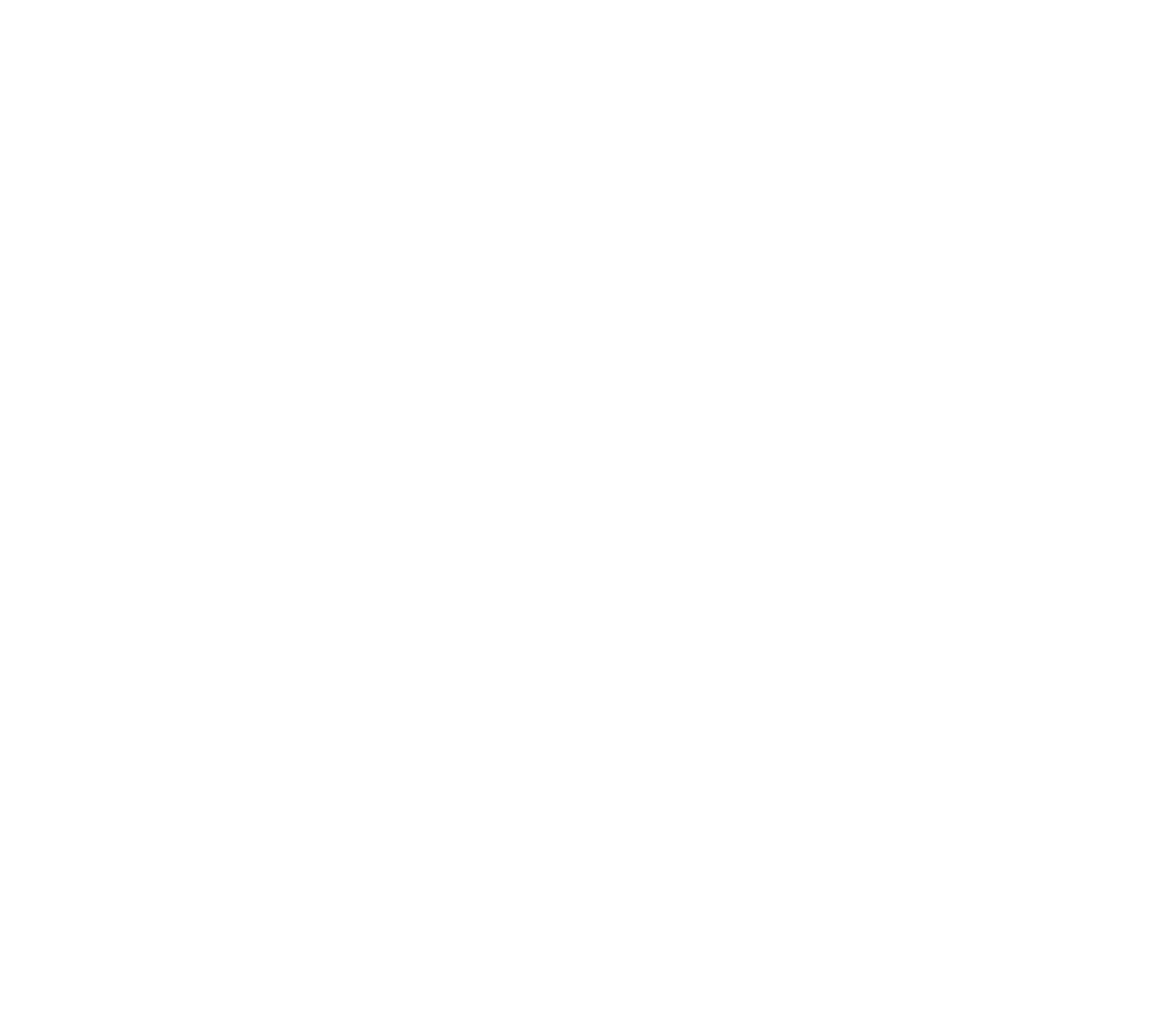}
    \put(0, 75){\includegraphics[width=0.25\linewidth]{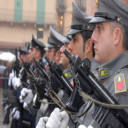}}
    \put(25, 75){\includegraphics[width=0.25\linewidth]{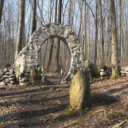}}
    \put(50, 75){\includegraphics[width=0.25\linewidth]{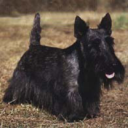}}
    \put(75, 75){\includegraphics[width=0.25\linewidth]{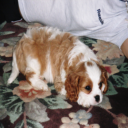}}

    \put(0, 50){\includegraphics[width=0.25\linewidth]{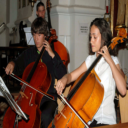}}
    \put(25, 50){\includegraphics[width=0.25\linewidth]{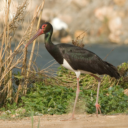}}
    \put(50, 50){\includegraphics[width=0.25\linewidth]{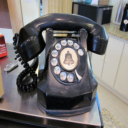}}
    \put(75, 50){\includegraphics[width=0.25\linewidth]{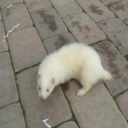}}

    \put(0, 25){\includegraphics[width=0.25\linewidth]{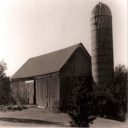}}
    \put(25, 25){\includegraphics[width=0.25\linewidth]{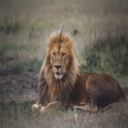}}
    \put(50, 25){\includegraphics[width=0.25\linewidth]{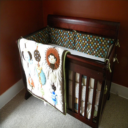}}
    \put(75, 25){\includegraphics[width=0.25\linewidth]{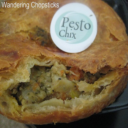}}
    
    \put(0, 0){\includegraphics[width=0.25\linewidth]{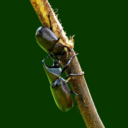}}
    \put(25, 0){\includegraphics[width=0.25\linewidth]{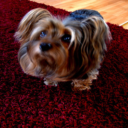}}
    \put(50, 0){\includegraphics[width=0.25\linewidth]{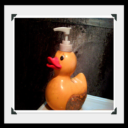}}
    \put(75, 0){\includegraphics[width=0.25\linewidth]{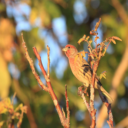}}
    \end{overpic}
    }
    \subcaptionbox{\label{subfig:recon_fullbatch}}[0.495\linewidth]{
        \begin{overpic}[width=\linewidth, height=\linewidth]{figs/4x4.pdf}
        \put(0, 75){\includegraphics[width=0.25\linewidth]{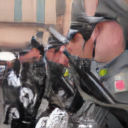}}
        \put(25, 75){\includegraphics[width=0.25\linewidth]{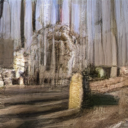}}
        \put(50, 75){\includegraphics[width=0.25\linewidth]{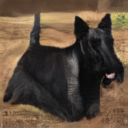}}
        \put(75, 75){\includegraphics[width=0.25\linewidth]{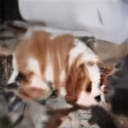}}
    
        \put(0, 50){\includegraphics[width=0.25\linewidth]{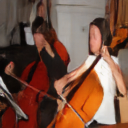}}
        \put(25, 50){\includegraphics[width=0.25\linewidth]{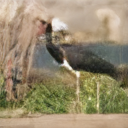}}
        \put(50, 50){\includegraphics[width=0.25\linewidth]{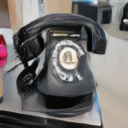}}
        \put(75, 50){\includegraphics[width=0.25\linewidth]{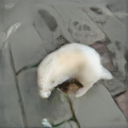}}
    
        \put(0, 25){\includegraphics[width=0.25\linewidth]{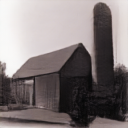}}
        \put(25, 25){\includegraphics[width=0.25\linewidth]{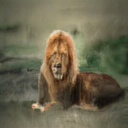}}
        \put(50, 25){\includegraphics[width=0.25\linewidth]{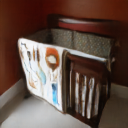}}
        \put(75, 25){\includegraphics[width=0.25\linewidth]{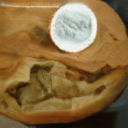}}
        
        \put(0, 0){\includegraphics[width=0.25\linewidth]{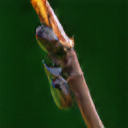}}
        \put(25, 0){\includegraphics[width=0.25\linewidth]{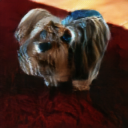}}
        \put(50, 0){\includegraphics[width=0.25\linewidth]{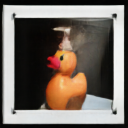}}
        \put(75, 0){\includegraphics[width=0.25\linewidth]{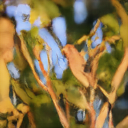}}
        \end{overpic}
        }

    \caption{
    Reconstruction results on a whole batch of 16 when raw gradients are known.
    Compared with (a)~the original images, (b)~the corresponding ROG reconstructed images are visually similar.
    \label{fig:full_batch}}
\end{figure}

Three different model architectures are considered as the instances of shared model in federated learning, namely, LeNet~\cite{zhu2019deep}, VGG-7~\cite{simonyan2015very}, and ResNet-18~\cite{he2016deep}.
As the batch normalization layers are reported to cause accuracy drop and raise privacy concerns in federated learning~\cite{hsieh2020non, huang2021evaluating}, we remove these layers in our study and leave them for future work. 
The attack results that focused on the batch normalization regularizer~\cite{yin2021see} are not included. 
For the reconstruction attack, we use the ImageNet validation dataset as the client private data. 
All images are resized to the resolution of $128 \times 128$. 
We set the batch size to $B=16$ and the number of local epochs to $\tau=5$. 
The learning rate is set to $5 \times 10^{-3}$.
We keep these settings throughout the paper unless otherwise specified.
We randomly select four reconstruction results and compare them with InvertGrad~\cite{geiping2020inverting}. 
We report average LPIPS (defined in Appendix~\ref{app:IQA}) of the whole batch.
Because we observe consistent reconstruction results across three neural network architectures, we present only the LeNet attack results in Figure~\ref{fig:attack_fedavg} and in subsequent experiments. 
Figure~\ref{fig:full_batch} shows the reconstructed images in the full batch, and we compare the best and the worst reconstructed images in Appendix~\ref{section:more_example}. 
The reconstruction on other neural network architectures can be found in Appendix~\ref{section:add_results}.
ROG reconstruction results are visually similar to the original image $\X_{m,i}$'s. 
In comparison, InvertGrad fails to reconstruct private image data. 

\highlight{Comparison with InvertGrad.}
The main differences between ROG and InvertGrad are twofold.
First, we propose to find the numerical solution to the optimization problem of \eqref{eq:new_optim_problem} in a low dimensional space. 
Compared to InvertGrad that directly looks for the solution in the original image space, ROG converges faster and yields more stable reconstruction results (see ``Ablation study'' below). 
Second, InvertGrad employs the total variation loss $\mathcal{L}_{\text{TV}}$ as an image prior.   
In our study, we train a neural network model on an additional public image dataset as a postprocessing module, which can be considered as a stronger image prior compared to the total variation regularization term $\mathcal{L}_{\text{TV}}$.  
Note that InvertGrad fails to reconstruct the data based on the raw gradients, and we do not include their attack results in the subsequent empirical studies. 
Please refer to Appendix~\ref{section:add_results} for the detailed comparison with InvertGrad/DLG on a whole batch of $16$.


\begin{figure}[tb]
    \centering
    \includegraphics[width=0.48\textwidth]{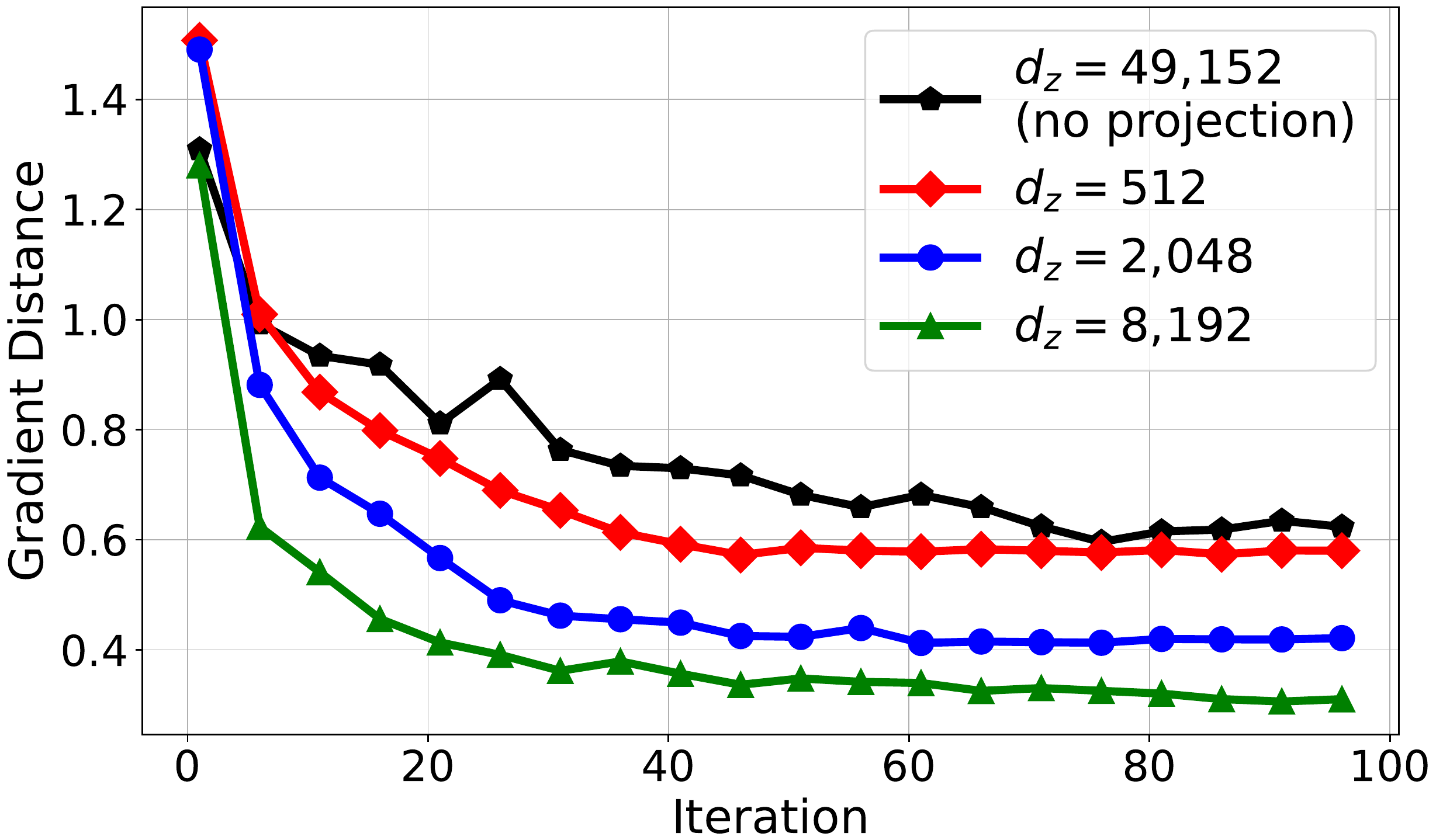}
    \caption{
    Gradient distance versus the optimization iteration when changing the dimension $d_z$ of the vectorization of the latent representation $\Z_{m,i}$.
    The three schemes based on latent weight projection converge faster than the baseline ``No Projection'', which can be treated as using identity functions for $\enc(\cdot)$ and $\dec(\cdot)$.  
    When $d_z$ further decreases, the convergence becomes slower.  
    }
    \label{fig:effect_dim}
\end{figure}

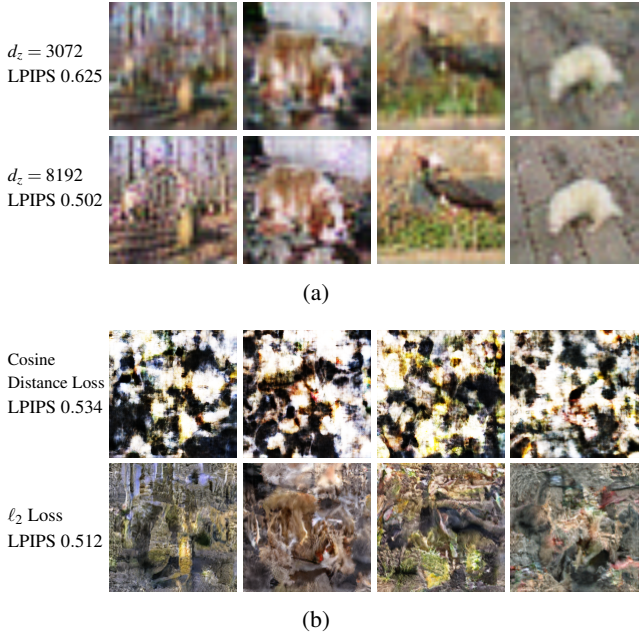
\begin{figure}[!tb]
    \vspace*{-6.6pt}
    \subcaptionbox{\label{subfig:without_postprocessing}}{
        \begin{overpic}[width=\linewidth, height=0.45\linewidth]{figs/4x4.pdf}
        \put(17, 21){\includegraphics[width=0.2\linewidth]{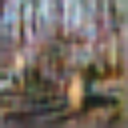}}
        \put(38, 21){\includegraphics[width=0.2\linewidth]{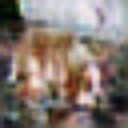}}
        \put(59, 21){\includegraphics[width=0.2\linewidth]{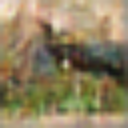}}
        \put(80, 21){\includegraphics[width=0.2\linewidth]{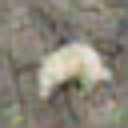}}

        \put(17, 0){\includegraphics[width=0.2\linewidth]{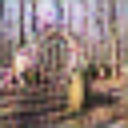}}
        \put(38, 0){\includegraphics[width=0.2\linewidth]{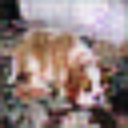}}
        \put(59, 0){\includegraphics[width=0.2\linewidth]{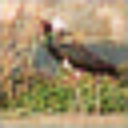}}
        \put(80, 0){\includegraphics[width=0.2\linewidth]{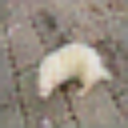}}

        \put(0, 32.5){\zoom{$d_z = 3072$}}
        \put(0, 28.5){\zoom{LPIPS $0.625$}}
        \put(0, 13){\zoom{$d_z = 8192$}}
        \put(0, 9){\zoom{LPIPS $0.502$}}
        \end{overpic}
    }
    \subcaptionbox{\label{subfig:without_latent}}{
    \begin{overpic}[width=\linewidth, height=0.45\linewidth]{figs/4x4.pdf}
        \put(17, 21){\includegraphics[width=0.2\linewidth]{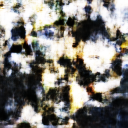}}
        \put(38, 21){\includegraphics[width=0.2\linewidth]{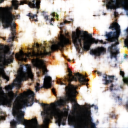}}
        \put(59, 21){\includegraphics[width=0.2\linewidth]{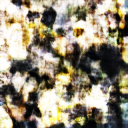}}
        \put(80, 21){\includegraphics[width=0.2\linewidth]{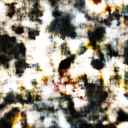}}
    
        \put(17, 0){\includegraphics[width=0.2\linewidth]{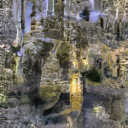}}
        \put(38, 0){\includegraphics[width=0.2\linewidth]{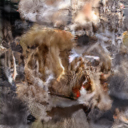}}
        \put(59, 0){\includegraphics[width=0.2\linewidth]{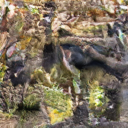}}
        \put(80, 0){\includegraphics[width=0.2\linewidth]{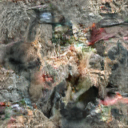}}

        \put(0, 35.5){\zoom[0.65]{Cosine }}
        \put(0, 32){\zoom[0.65]{Distance Loss}}
        \put(0, 28){\zoom{LPIPS $0.534$}}
        \put(0, 11){\zoom{$\ell_2$ Loss}}
        \put(0, 7){\zoom{LPIPS $0.512$}}
    \end{overpic}
    }
    \vspace*{-5pt}
    \caption{
    Ablation study of (a)~reconstruction without postprocessing ($100$ iterations) and 
    (b)~reconstruction without using low-dimensional representation ($20$k iterations).  
    Two modules work synergistically and do not perform well individually.
    \label{fig:ablation}}
\end{figure}

\highlight{Ablation Study. }
We further investigate the effect of two components in ROG attack, including low-dimensional representation and image postprocessing. 
\textbf{First}, we remove the postprocessing module and compare the convergence rate under different dimensions of representations $\Z_{m,i}$.
We optimize three autoencoders~\cite{baldi2012autoencoders} to obtain different encoding functions $\enc(\cdot)$ and the decoding functions $\dec(\cdot)$ in ROG attack. 
For autoencoders, we set the dimension $d_z$ of the vectorization of the representation $\Z_{m,i}$ to $8192$, $2048$, and $512$, respectively. 
In the reconstruction attack, the dimension $d_z$ is equal to the number of unknown variables that an adversary wants to reconstruct. 
A smaller $d_z$ makes it easier for the attacker to solve the system, but will also lead to more information loss during the encoding and decoding procedure. 
For the baseline method without projection, the number of unknown variables is equal to the number of pixels in the images. 
In our study, where each RGB image has a resolution of $128\times 128$, the number of unknowns is $49,\!152$.  
The ``no projection'' baseline sets encoding function $\enc(\cdot)$ and decoding function $\dec(\cdot)$ to identity functions.
The optimization curves of different attack schemes are shown in Figure~\ref{fig:effect_dim}. 
It can be observed that the three methods with the autoencoder projection converge faster than the baseline method without projection. 
The results confirm the effectiveness of the latent vector projection in ROG.  
When the dimension $d_z$ further decreases, the convergence is slower. 
We visualize the reconstruction when using bicubic downsampled representation ($d_z = 3072$) and autoencoder latent representation ($d_z = 8192$) in \paperfig{subfig:without_postprocessing}. 
The number of iterations is set to $100$. 
The autoencoder representation ($d_z = 8192$) gives slightly better results. 
Overall, reconstructed images are distorted and blurry with certain structural information revealed. 
\textbf{Second}, we remove the low-dimensional projection step and apply the postprocessing module directly. 
This is equivalent to further postprocess the output of InvertGrad/DLG.
The number of iterations is set to $20$k, following~\cite{geiping2020inverting,yin2021see}. 
The results are shown in \paperfig{subfig:without_latent}, which are unrecognizable (raw images are in \paperfig{fig:attack_fedavg}). 
Combing the two components, ROG improves efficiency and reconstruction quality. 
In particular, the number of iterations can be reduced from $10^{4}$ to $10^{2}$. 
The quality improvement is more than $70\%$, improving LPIPS from $0.7$ to $0.2$. 
From the two experiments above, we conclude that the low-dimensional representation and postprocessing are both necessary to ensure fast and high-quality reconstruction.

\subsection{Attack on Compressed Gradients}

\begin{figure}[!tb]
    \begin{overpic}[width=\linewidth]{figs/4x4.pdf}
    \put(17, 69){\includegraphics[width=0.2\linewidth]{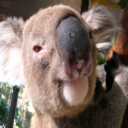}}
    \put(38, 69){\includegraphics[width=0.2\linewidth]{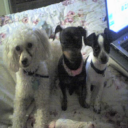}}
    \put(59, 69){\includegraphics[width=0.2\linewidth]{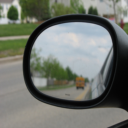}}
    \put(80, 69){\includegraphics[width=0.2\linewidth]{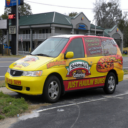}}

    \put(17, 67){\includegraphics[width=0.83\linewidth, height=1.3pt]{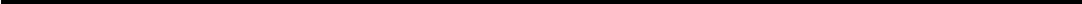}}

    \put(17, 45){\includegraphics[width=0.2\linewidth]{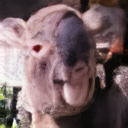}}
    \put(38, 45){\includegraphics[width=0.2\linewidth]{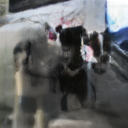}}
    \put(59, 45){\includegraphics[width=0.2\linewidth]{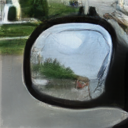}}
    \put(80, 45){\includegraphics[width=0.2\linewidth]{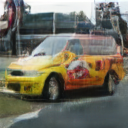}}

    \put(17, 24){\includegraphics[width=0.2\linewidth]{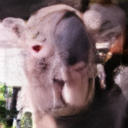}}
    \put(38, 24){\includegraphics[width=0.2\linewidth]{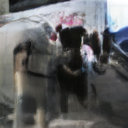}}
    \put(59, 24){\includegraphics[width=0.2\linewidth]{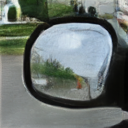}}
    \put(80, 24){\includegraphics[width=0.2\linewidth]{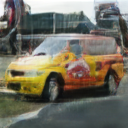}}

    \put(17, 3){\includegraphics[width=0.2\linewidth]{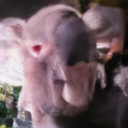}}
    \put(38, 3){\includegraphics[width=0.2\linewidth]{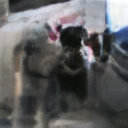}}
    \put(59, 3){\includegraphics[width=0.2\linewidth]{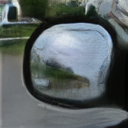}}
    \put(80, 3){\includegraphics[width=0.2\linewidth]{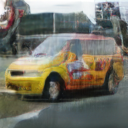}}

    \put(0, 77){\zoom{Raw Images}}
    \put(0, 55){\zoom{$3$-bit $\pfunc_{\text{q}}$}}
    \put(0, 51){\zoom{LPIPS $0.213$}}
    \put(0, 34.5){\zoom{$3$-bit $\pfunc_{\text{qsgd}}$}}
    \put(0, 30.5){\zoom{LPIPS $0.215$}}
    \put(0, 15){\zoom{Top-$k$ $(0.95)$}}
    \put(0, 11){\zoom{LPIPS $0.237$}}
    \end{overpic}
    \vspace*{-25pt}
    \caption{Reconstructed images when gradient compression is applied in FedAvg. 
    The compression methods include: $3$-bit uniform quantizer $\pfunc_{\text{q}}$, $3$-bit QSGD quantizer $\pfunc_{\text{qsgd}}$, and Top-$k$ method $\pfunc_{\text{topk}}$ with the sparsity set to $0.95$.   
    The reconstructed images are visually similar to the raw private images. 
    \label{fig:attack_gradcomp}}
\end{figure}
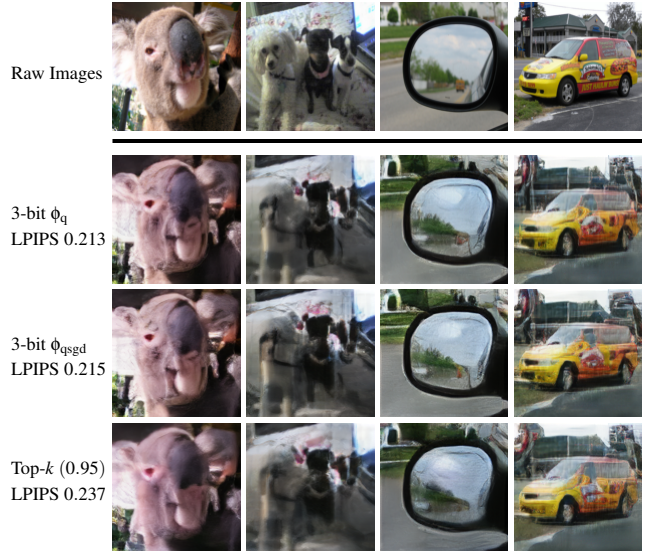

We now discuss the attack results when the compressed gradients are transmitted. 
The experimental setups are the same as Section~\ref{section:realization}. 
We use a $3$-bit uniform quantizer $\pfunc_{\text{q}}$, a $3$-bit QSGD quantizer $\pfunc_{\text{qsgd}}$, and the Top-$k$ method $\pfunc_{\text{topk}}$ with the sparsity parameter equal to $0.95$. 
The reconstruction results are shown in Figure~\ref{fig:attack_gradcomp}.
It can be observed that our reconstruction has revealed visual information of the original image data, indicating that even the low precision quantization and sparsification schemes do not provide privacy enhancement despite the information loss. 
Contrary to prior empirical studies~\cite{zhu2019deep,wei2020framework}, we demonstrate that gradient compression should not be automatically treated as a defense scheme against reconstruction attacks. 
Prior work on communication-efficient federated learning showed that gradient compression does not significantly affect the model accuracy~\cite{alistarh2017qsgd, aji2017sparse, reisizadeh2020fedpaq}. 
From the perspective of the ROG reconstruction, we confirm the intuition that compressed gradients still preserve most of the information with respect to the raw data. 
As gradient compression does not incorporate privacy protection by design, the observations of failed reconstruction in the literature may give a false sense of security.  

\begin{figure}[!tb]
    \begin{overpic}[width=\linewidth]{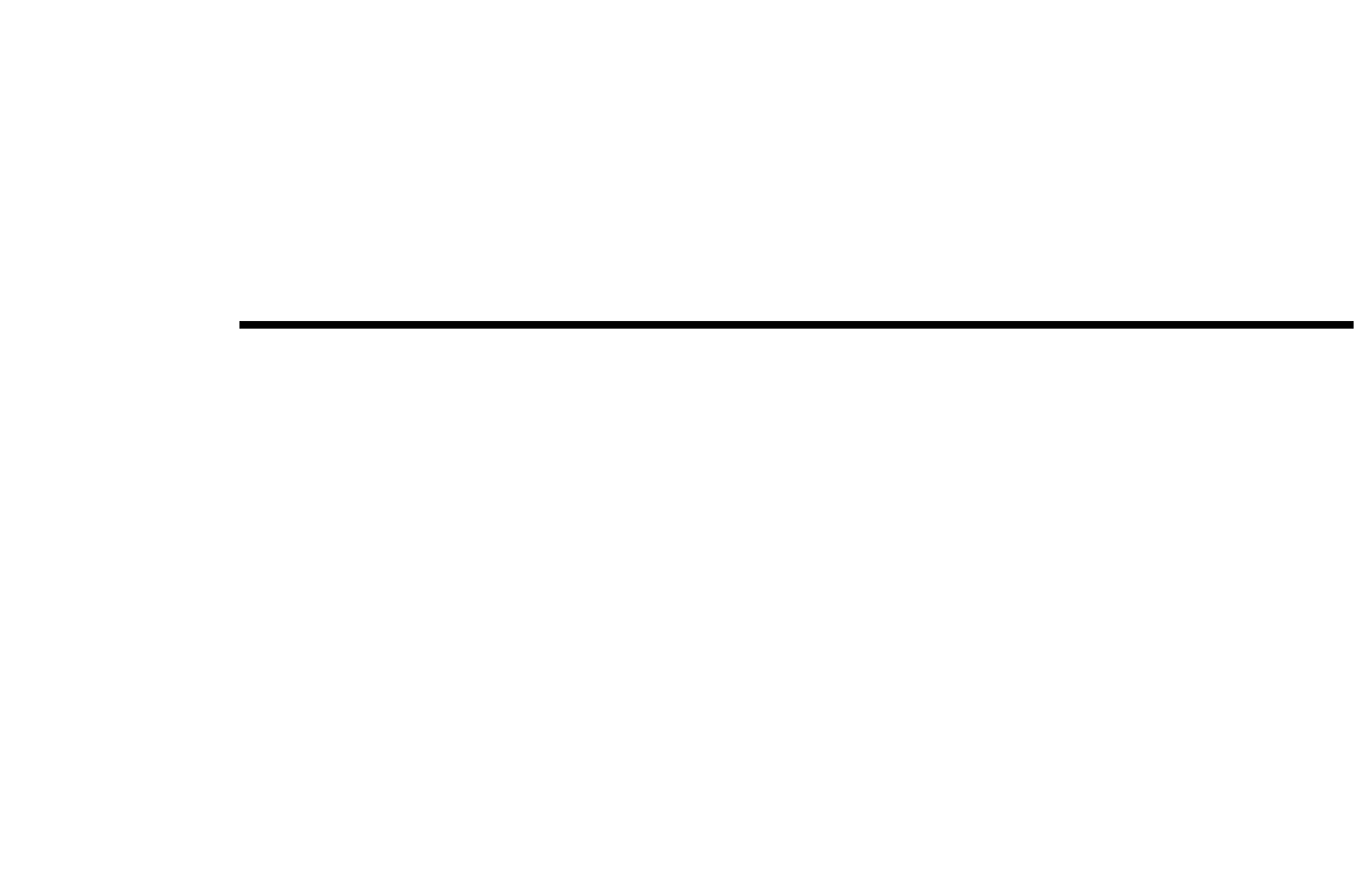}
    \put(17, 46){\includegraphics[width=0.2\linewidth]{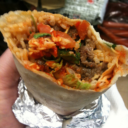}}
    \put(38, 46){\includegraphics[width=0.2\linewidth]{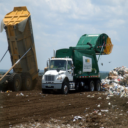}}
    \put(59, 46){\includegraphics[width=0.2\linewidth]{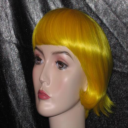}}
    \put(80, 46){\includegraphics[width=0.2\linewidth]{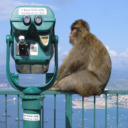}}

    \put(17, 44){\includegraphics[width=0.825\linewidth, height=1.3pt]{figs/lines/line.png}}

    \put(17, 22.){\includegraphics[width=0.2\linewidth]{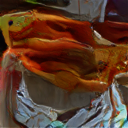}}
    \put(38, 22.){\includegraphics[width=0.2\linewidth]{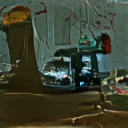}}
    \put(59, 22.){\includegraphics[width=0.2\linewidth]{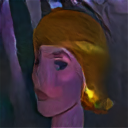}}
    \put(80, 22.){\includegraphics[width=0.2\linewidth]{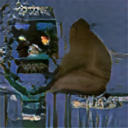}}

    \put(17, 1){\includegraphics[width=0.2\linewidth]{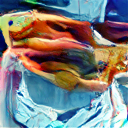}}
    \put(38, 1){\includegraphics[width=0.2\linewidth]{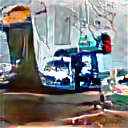}}
    \put(59, 1){\includegraphics[width=0.2\linewidth]{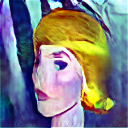}}
    \put(80, 1){\includegraphics[width=0.2\linewidth]{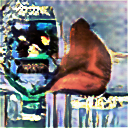}}

    \put(0, 54){\zoom{Raw Images}}
    \put(0, 32){\zoom{signSGD}}
    \put(0, 28){\zoom{LPIPS $0.347$}}
    
    \put(0, 14){\zoom[0.64]{Postprocessed}}
    \put(0, 10){\zoom[0.64]{by Histogram}}
    \put(0, 6){\zoom[0.64]{Equalization}}
    \end{overpic}
    \vspace*{-20pt}
    \caption{Reconstructed images from signSGD. 
    Despite the loss of magnitude information caused by 1-bit quantization, ROG can reconstruct results visually similar to the raw images.  
    \label{fig:attack_signsgd}}
\end{figure}

\highlight{Reconstruction from signSGD~\cite{bernstein2018signsgd}.}\label{section:signsgd}
The ROG attack method can be extended to the scenario when the $1$-bit compressor $\sign(\cdot)$ is adopted.  
Bernstein et al.~\cite{bernstein2018signsgd} showed that the signSGD algorithm converges fast even though only the signs of the gradients are transmitted. 
In our attack, we modify the original optimization problem \eqref{eq:new_optim_problem} to 
\begin{equation}\label{eq:optim_problem_signSGD}
    \Z^\star_{m,i} = \argmin_{\Z_{m,i}} \left\| \tanh(\grad') - \sign(\grad) \right\|^2, 
\end{equation}
where $\grad$ is the client gradient, $\grad'$ is the dummy gradient generated in the attack framework, and $\tanh(\cdot)$ is the hyperbolic tangent function approximating the $\sign(\cdot)$ operator. 
We use the differentiable $\tanh(\cdot)$ function to facilitate the SGD-based optimization of \eqref{eq:optim_problem_signSGD}. 
After the optimization terminates, we normalize each image with the maximum absolute values across all coordinates. 
The results are shown in Figure~\ref{fig:attack_signsgd}.
Since the $\sign(\cdot)$ operator does not preserve the magnitude information, the original normalized images on the second row tend to have a reduced contrast.
We further apply the histogram equalization algorithm to enhance the contrast~\cite{gonzalez2014digital}. 
Counterintuitively, although the $1$-bit $\sign(\cdot)$ operator is applied and a great amount of information is discarded, the reconstructed images reveal important information such as the dominant objects in the foreground and the structure information in the background.

\section{Attack on Existing Defenses}\label{section:current_defense}

In this section, we discuss the attack results under three state-of-the-art defense schemes, including Soteria~\cite{sun2021soteria}, PRECODE~\cite{scheliga2022precode}, and FedCDP~\cite{wei2021gradient}.
All defenses are reported to be effective against the existing gradient leakage attacks, including DLG and InvertGrad. 
For clarity, we first briefly review these schemes and then provide the empirical attack results. 

\subsection{Review of Defense Schemes}

\highlight{Soteria~\cite{sun2021soteria}.}
Soteria chooses a fully connected layer of the convolutional neural network model as a \textit{defended layer} and perturbs the data representation. 
Let $\X$ and $\X^{\prime}$ denote the raw image and reconstructed image via the perturbed representation, respectively. 
Their corresponding data representations in the defended layer are denoted as $\rep$ and $\rep'$. 
To reduce the risk of information leakage, Soteria formulates the problem by maximizing the distance between the original image $\X$ and the reconstructed version $\X'$, while maintaining the similarity between the original representation $\rep$ and the target representation $\rep'$.  
The constrained optimization problem is given as: 
\begin{subequations}
\begin{align}
    \max_{r^{\prime}} & \left\|\X-\X^{\prime}\right\|_{2}, \\
    \text { s.t. } & \left\|\rep-\rep^{\prime}\right\|_{0} \leqslant \rho.
\end{align}
\end{subequations}
In Soteria, the $\ell_0$ norm can ensure the sparsity and its effect can be equivalently represented by a pruning rate. 
Since this defense has an optimization stage, it will introduce non-negligible computational overhead, especially for the nodes in the fully-connected layers. 
It has been shown that Soteria is robust and does not impede the convergence of FedAvg~\cite{sun2021soteria}. 
However, this defense does not provide provable privacy protection.
For example, Balunovi{\'c}~\cite{balunovic2021bayesian} discussed that Soteria only introduced the randomness to the defended layer, and an attacker may still compromise the model by circumventing it.  
In our attack, we directly reconstruct the noisy image $\X'$ by using the perturbed representation and utilizing the postprocessing tools to enhance the image quality. 
We will demonstrate that the postprocessing network in the ROG framework is robust to the perturbation in the image space. 

\highlight{PRECODE~\cite{scheliga2022precode}.}
PRECODE perturbs the data representation by adding a variational block between two successive layers in a neural network. 
The variational block is composed of an encoder and a decoder, which are realized as fully connected layers.  
Given an input $\x$, the encoder projects $\x$ to a representation $\z$, and then generates a feature $\vecb \sim q(\vecb \mid \z)$ from a Gaussian distribution. 
The decoder reconstructs a representation $\hat{\z} = p(\z \mid \vecb) p(\vecb)$.  

The randomness comes from the sampling of the random vector in the variational block and will affect the gradients in other layers in the backpropagation.  
Compared to Soteria, an attacker cannot bypass this layer to remove the effect of perturbation. 
On the other hand, the variational block is available to the attacker based on a white-box model assumption.  
We let the attacker match the gradients in the variational block without modifying the ROG pipeline. 
We will show that PRECODE does not effectively protect federated learning against the data reconstruction attack. 

\highlight{FedCDP~\cite{wei2021gradient}.}
FedCDP is a gradient leakage resilient defense based on client level differential privacy (DP) mechanism. 
It applies per-example gradient clipping and noise injection to achieve provable DP guarantees. 
In each layer $\ell$, the gradient $\grad_{m,i,\ell}$ is computed for every individual training example $\X_{m,i}$, and the clipping step is formulated as 
\begin{equation}
    \hat{\grad}_{m,i,\ell} = \grad_{m,i,\ell}/\max\left(1, \frac{\|\grad_{m,i,\ell}\|_2}{C}\right),
\end{equation}
where $C$ is a constant of the clipping upper bound. 
The Gaussian noise vector is then added to the clipped gradient, namely,
\begin{equation}
    \pfunc_{\text{dp}}(\hat{\grad}) = \hat{\grad} + \mathbf{n},\quad n_i \sim \mathcal{N}(0, \sigma^2 C^2).
\end{equation}
Wei et al.~\cite{wei2021gradient} have shown that FedCDP can achieve client level per-example differential privacy, and the scheme is robust to data reconstruction attacks.  

However, differential privacy may not explicitly protect attribute privacy. 
For example, Zhang et al.~\cite{zhang2020secret} studied the model inversion problem in centralized learning. 
They found that the attack accuracy remains unchanged despite the various settings of differential privacy budgets. 
In this work, we empirically study the attack against the client differentially private algorithm in federated learning. 
Our results indicate that the popular differentially private gradient descent training~\cite{abadi2016deep} and the federated implementation~\cite{wei2021gradient} may need to be carefully redesigned.

\begin{figure}[!tb]
    \begin{overpic}[width=\linewidth]{figs/4x4.pdf}
    \put(17, 69){\includegraphics[width=0.2\linewidth]{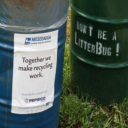}}
    \put(38, 69){\includegraphics[width=0.2\linewidth]{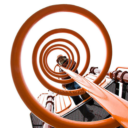}}
    \put(59, 69){\includegraphics[width=0.2\linewidth]{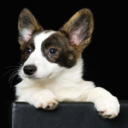}}
    \put(80, 69){\includegraphics[width=0.2\linewidth]{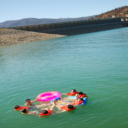}}

    \put(17, 67){\includegraphics[width=0.83\linewidth, height=1.3pt]{figs/lines/line.png}}

    \put(17, 45){\includegraphics[width=0.2\linewidth]{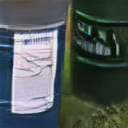}}
    \put(38, 45){\includegraphics[width=0.2\linewidth]{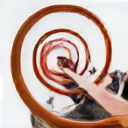}}
    \put(59, 45){\includegraphics[width=0.2\linewidth]{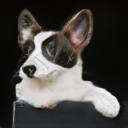}}
    \put(80, 45){\includegraphics[width=0.2\linewidth]{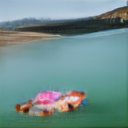}}

    \put(17, 24){\includegraphics[width=0.2\linewidth]{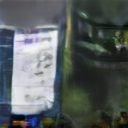}}
    \put(38, 24){\includegraphics[width=0.2\linewidth]{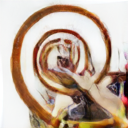}}
    \put(59, 24){\includegraphics[width=0.2\linewidth]{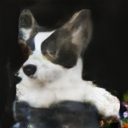}}
    \put(80, 24){\includegraphics[width=0.2\linewidth]{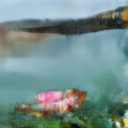}}

    \put(17, 3){\includegraphics[width=0.2\linewidth]{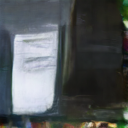}}
    \put(38, 3){\includegraphics[width=0.2\linewidth]{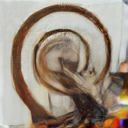}}
    \put(59, 3){\includegraphics[width=0.2\linewidth]{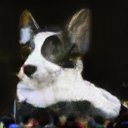}}
    \put(80, 3){\includegraphics[width=0.2\linewidth]{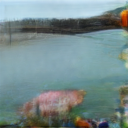}}

    \put(0, 76){\zoom{Raw Images}}
    \put(0, 55){\zoom{Soteria}}
    \put(0, 51){\zoom{LPIPS $0.180$}}
    \put(0, 35.5){\zoom{PRECODE}}
    \put(0, 31.5){\zoom{LPIPS $0.307$}}
    \put(0, 16){\zoom[0.59]{FedCDP ($0$ dB)}}
    \put(0, 12){\zoom{LPIPS $0.376$}}
    \end{overpic}
    \vspace*{-20pt}
    \caption{
    Attack against defense schemes Soteria~\cite{sun2021soteria}, PRECODE~\cite{scheliga2022precode}, and FedCDP~\cite{wei2021gradient}. 
    The two feature perturbation based schemes, Soteria and PRECODE, are not effective to defend against the ROG reconstruction attack. 
    Furthermore, FedCDP may not protect attribute privacy effectively.  
    It can be observed that most of the structural information can be reconstructed. 
    \label{fig:attack_defense}}
\end{figure}

\subsection{Attack Results}

\begin{figure*}[!tb]
    \centering
    \subcaptionbox{\label{subfig:tradeoff}}[0.56\linewidth]{
    \begin{overpic}[width=\linewidth]{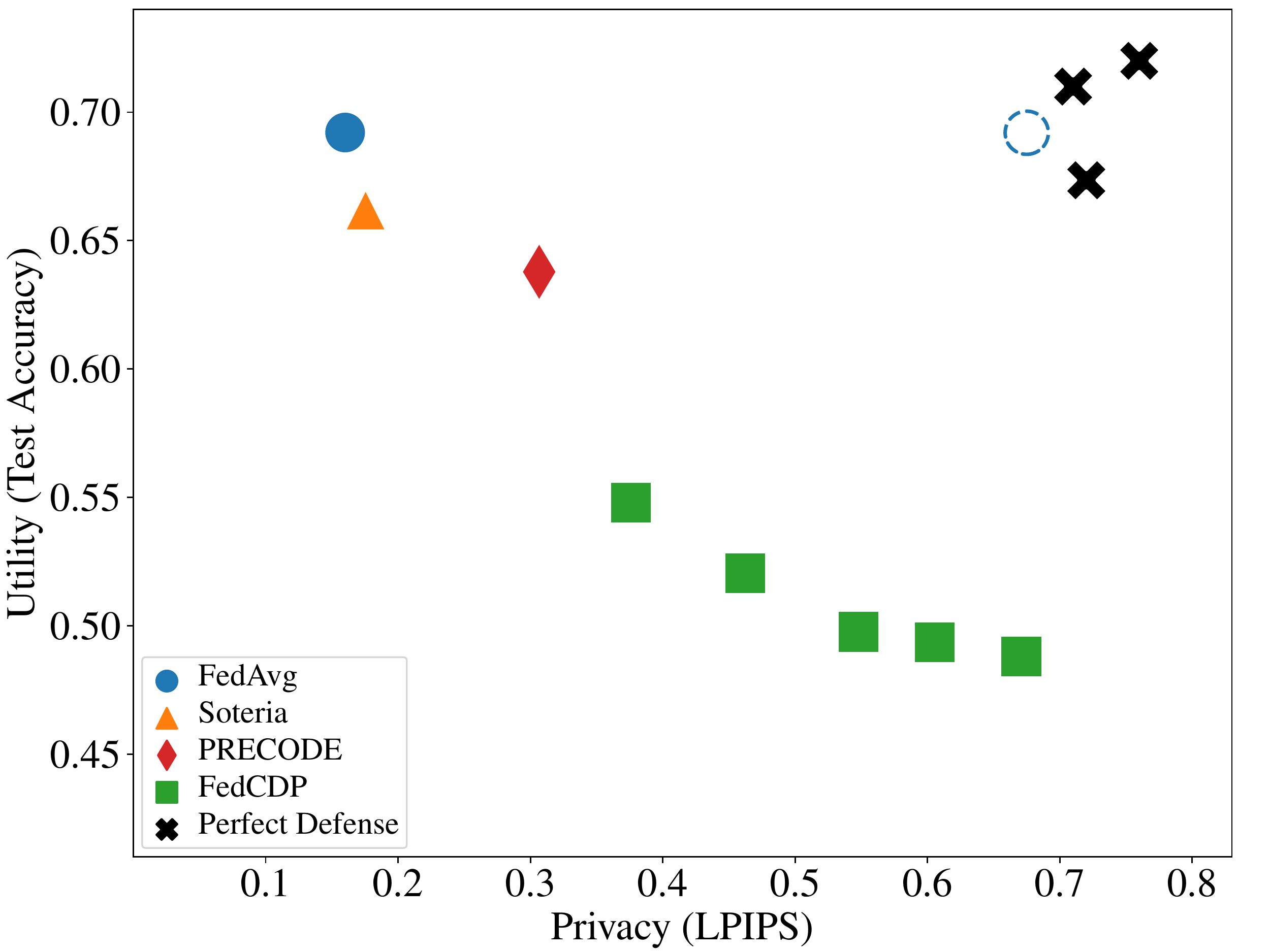}
        \put(14.5, 41){\includegraphics[width=0.15\linewidth]{figs/defense/4_soteria.png}}
        \put(28, 35.5){\includegraphics[width=0.15\linewidth]{figs/defense/4_precode.png}}

        \put(36, 18.5){\includegraphics[width=0.15\linewidth]{figs/defense/4_fedcdp.png}}
        \put(45, 13.){\includegraphics[width=0.15\linewidth]{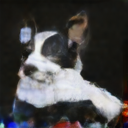}}
        \put(54, 8.5){\includegraphics[width=0.15\linewidth]{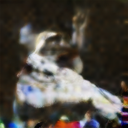}}
        \put(63, 51){\includegraphics[width=0.15\linewidth]{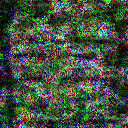}}
        
        \put(11, 72){\intab{Achievable Utility-Privacy Region}}
        \put(18, 69){\intab{(Proposed: ROG)}}

        \put(51, 72){\intab{Achievable Utility-Privacy Region}}
        \put(58, 69){\intab{(Prior: InvertGrad)}}

        \put(76, 45){\intab{Improved}}
        \put(76, 41){\intab{Attack Efficiency}}
        \put(76, 37){\intab{by ROG}}
        \put(18.2, 9){\includegraphics[width=0.82\linewidth]{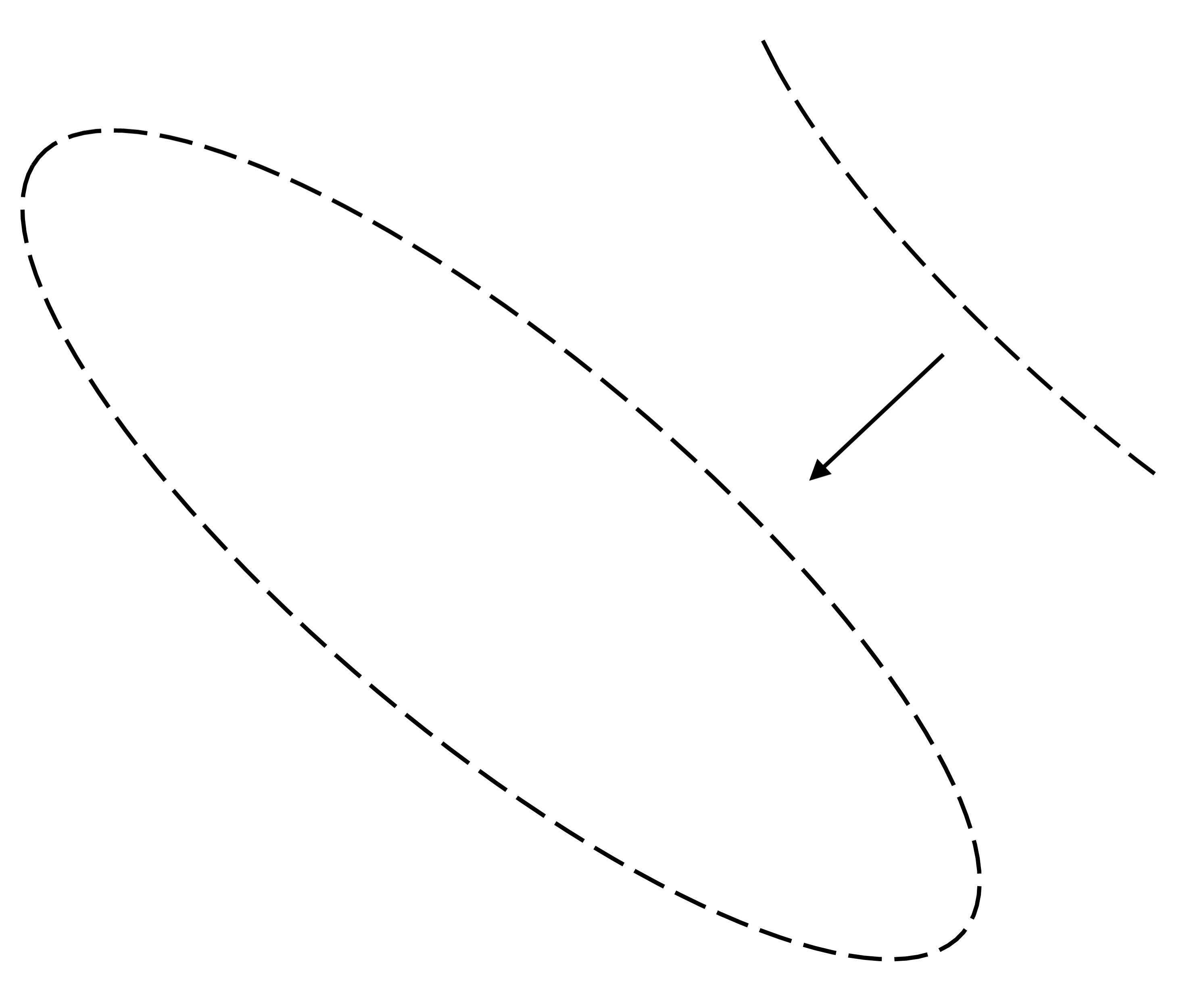}}
    
        \put(47, 38){\intab{0 dB}}
        \put(56, 32.5){\intab{-5 dB}}
        \put(63, 28.5){\intab{-10 dB}}
        \put(69, 26.5){\intab{-15 dB}}
        \put(76, 25.5){\intab{-20 dB}}
    \end{overpic}}\hfill
    \subcaptionbox{\label{subfig:fedcdp_qualitative}}[0.4\linewidth]{
    \begin{overpic}[width=\linewidth]{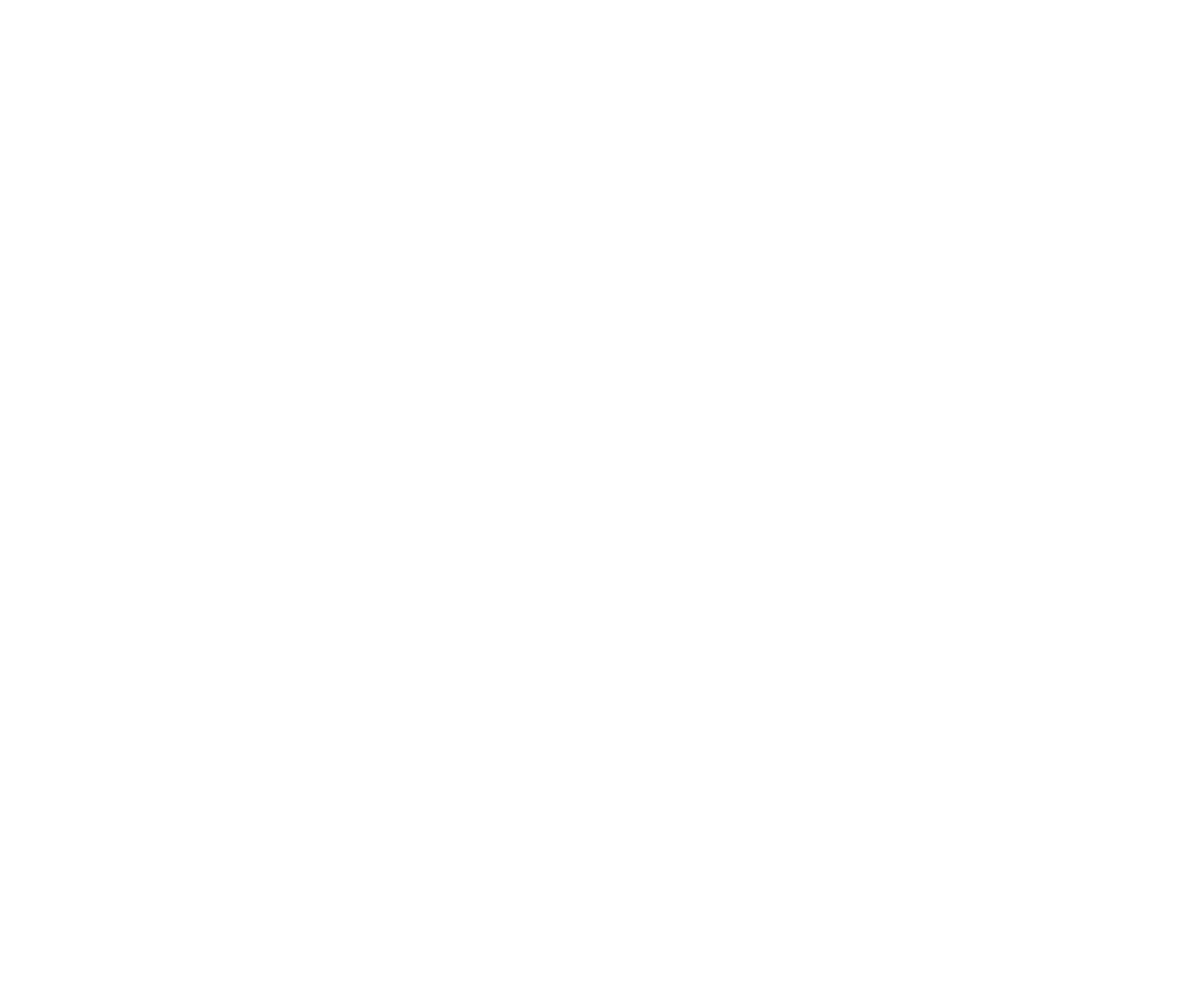}
        \put(17, 85){\includegraphics[width=0.2\linewidth]{figs/defense/1_ori.png}}
        \put(38, 85){\includegraphics[width=0.2\linewidth]{figs/defense/2_ori.png}}
        \put(59, 85){\includegraphics[width=0.2\linewidth]{figs/defense/4_ori.png}}
        \put(80, 85){\includegraphics[width=0.2\linewidth]{figs/defense/12_ori.png}}
    
        \put(17, 83){\includegraphics[width=0.826\linewidth, height=1.3pt]{figs/lines/line.png}}

        \put(17, 61.5){\includegraphics[width=0.2\linewidth]{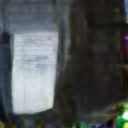}}
        \put(38, 61.5){\includegraphics[width=0.2\linewidth]{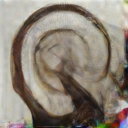}}
        \put(59, 61.5){\includegraphics[width=0.2\linewidth]{figs/fedcdp/_5/4_recon.png}}
        \put(80, 61.5){\includegraphics[width=0.2\linewidth]{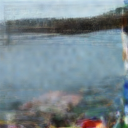}}
    
        \put(17, 41){\includegraphics[width=0.2\linewidth]{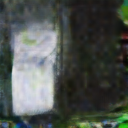}}
        \put(38, 41){\includegraphics[width=0.2\linewidth]{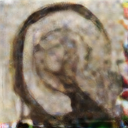}}
        \put(59, 41){\includegraphics[width=0.2\linewidth]{figs/fedcdp/_10/4_recon.png}}
        \put(80, 41){\includegraphics[width=0.2\linewidth]{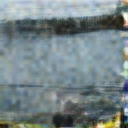}}
    
        \put(17, 20.5){\includegraphics[width=0.2\linewidth]{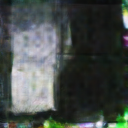}}
        \put(38, 20.5){\includegraphics[width=0.2\linewidth]{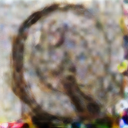}}
        \put(59, 20.5){\includegraphics[width=0.2\linewidth]{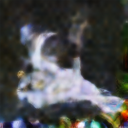}}
        \put(80, 20.5){\includegraphics[width=0.2\linewidth]{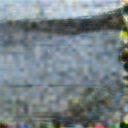}}
    
        \put(17, 0){\includegraphics[width=0.2\linewidth]{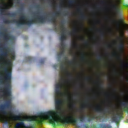}}
        \put(38, 0){\includegraphics[width=0.2\linewidth]{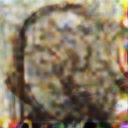}}
        \put(59, 0){\includegraphics[width=0.2\linewidth]{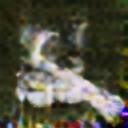}}
        \put(80, 0){\includegraphics[width=0.2\linewidth]{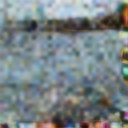}}

        \put(-4, 93){\zoom{Raw Images}}

        \put(-4, 70){\zoom{SNR $-\,5$ dB}}
        \put(-1, 74){\zoom{$\epsilon$ $\quad$ $15.2$}}
        \put(-4, 66){\zoom{LPIPS $0.462$}}
        
        \put(-4, 50){\zoom{SNR $-10$ dB}}
        \put(0, 54){\zoom{$\epsilon$ $\quad$ $4.6$}}
        \put(-4, 46){\zoom{LPIPS $0.548$}}
        
        \put(-4, 30){\zoom{SNR $-15$ dB}}
        \put(0, 34){\zoom{$\epsilon$ $\quad$ $1.9$}}
        \put(-4, 26){\zoom{LPIPS $0.605$}}

        \put(-4, 10){\zoom{SNR $-20$ dB}}
        \put(0, 14){\zoom{$\epsilon$ $\quad$ $1.0$}}
        \put(-4, 6){\zoom{LPIPS $0.671$}}
    \end{overpic}
    }
    \caption{
    (a)~Trade-off between model accuracy and privacy under the ROG attack. 
    A larger LPIPS value indicates better privacy protection against the attack.  
    The cross points in the upper-right corner represent the desired/unrealistic defense schemes achieving high model accuracy and privacy protection. 
    The dashed circle denotes the operating point of FedAvg under InvertGrad attack, which is misleadingly classified to the same region as the desired defense schemes and gives a false sense of security.  
    Meanwhile, the three defense schemes have been verified to be effective against the InvertGrad attack, which may also be classified as desired defense schemes. 
    The achievable utility--privacy region is shifted to the bottom left due to the proposed ROG attack. 
    (b)~ROG attack results when FedCDP is used as the defense scheme. 
    The reconstructed images gradually degrade as the defense strength increases (or the SNR decreases). 
    When the SNR is $-20$ dB, some structural information can still be revealed whereas the accuracy has dropped severely compared to FedAvg. 
    }
\end{figure*}

We first perform the ROG attack against the three defense schemes, Soteria, PRECODE, and FedCDP.
We use the same experimental setups as in Section~\ref{section:realization}.  
For Soteria, we follow Sun et al.~\cite{sun2021soteria} and use the configuration of the original paper to achieve a pruning rate of $80\%$ that provides the highest protection level.  
For FedCDP, we use a clipping upper bound $C=4$ adopted by Wei et al.~\cite{wei2021gradient}. 
We treat the clipped gradient $\hat{\grad}$ as the raw signal and use the signal-to-noise ratio (SNR) as the noise strength measurement. 
The results are shown in Figure~\ref{fig:attack_defense}.
We set the SNR to $0$ dB in FedCDP. 
The effect of the noise strength is discussed in the next experiment. 
We observe that the data representation based defenses, including Soteria and PRECODE, can still be vulnerable to the data reconstruction attack.  
Despite artifacts introduced in the postprocessing stage, the reconstructed results are close to the original images. 
In the meantime, FedCDP does not protect privacy effectively as the structural information and semantics can be revealed.
We provide more discussions as follows. 

\highlight{Trade-off between accuracy and privacy.}
In this experiment, we examine the trade-off between the model test accuracy and privacy leakage. 
We choose the SNR values from $\{-20, -15, -10, -5, 0\}$ dB for FedCDP, corresponds to $\epsilon \in \{ 108.0, 15.2, 4.6, 1.9, 1.0\}$.
To quantify privacy leakage, we perform the ROG attack targeting the ImageNet validation dataset assigned to clients. 
To measure the model accuracy, we choose the CIFAR-10 dataset~\cite{krizhevsky2009learning} with $c = 10$ categories as the image classification task in federated learning. 
We use a different dataset as the original ImageNet training data has been exploited during the postprocessing network training. 
Following Hsu et al.~\cite{hsu2019measuring}, we simulate the non-IID data with the symmetric Dirichlet distribution \cite{good1976application}. 
Specifically, for the $m$th client, we draw a random vector $\boldsymbol{q}_m \sim \text{Dir}(\alpha)$, where $\boldsymbol{q}_m = [q_{m,1}, \dots, q_{m,c}]^{\top}$ belongs to the $(c-1)$-standard simplex.  
Images with category $k$ are assigned to the $m$th client in proportional to $(100 \cdot q_{m,k})\%$. 
The parameter $\alpha$ is set to $0.5$.
A total of $100$ clients are considered, and $10$ of them are selected with equal probability in each communication round.  
We terminate the training after $100$ communication rounds and use the model accuracy as the utility metric. 
The trade-off between the model accuracy and privacy leakage of different methods is shown in Figure~\ref{subfig:tradeoff}. 
In general, defense schemes sacrifice the model accuracy to reduce privacy leakage. 
The cross points in the upper right corner represent desired defense schemes that achieve high model accuracy and privacy protection.
The proposed ROG attack shifts the achievable utility--privacy region to the bottom left, indicating that existing privacy-preserving federated learning algorithms may give a false sense of security.   
FedAvg and the three defense schemes can be effective against the InvertGrad attack, which may be misleadingly classified as desirable defense schemes. 
We present qualitative reconstruction results under different FedCDP settings in Figure~\ref{subfig:fedcdp_qualitative}.
When the SNR decreases, the reconstructed results gradually degrade.
Even when the SNR is low, some structural information can still be revealed. 
We note that there exists a relatively smooth transition for the attack results against different defense settings. 
As a result, it is ambiguous to tell whether an attack is successful or not.

Based on the observation, we point out that existing federated learning algorithms and defenses can still be vulnerable to reconstruction attacks. 
Differential privacy tools may not explicitly protect data attributes, such as the pixels of images in our simulation, while they could sacrifice the model utility to a large extent.  
Meanwhile, other defense schemes that do not rely on the differential privacy concept were evaluated on some specific attack designs without provable privacy guarantees.  
We have shown that these heuristic defenses do not provide satisfactory privacy protection.
Our results indicate the importance of redesigning the privacy-preserving framework for federated learning in parallel with the existing differential privacy paradigm.

\highlight{Unevaluated defenses.} 
There exist other defense schemes that we do not evaluate in this paper. 
For example, Huang et al.~\cite{huang2020instahide, huang2021evaluating} proposed to encode the images by compositing several training examples and flipping the signs of the data randomly. 
The vulnerability of data encoding has been shown in recent work~\cite{carlini2021private}: with a well-designed attack that formulated the problem as a noisy linear system, the original images can be recovered. 
Other defenses combine differential privacy with secure aggregation~\cite{agarwal2018cpsgd, kairouz2021distributed} to achieve stronger protection. 
Some studies have shown that the individual gradient can be recovered in secure aggregation~\cite{lam2021gradient}. 
A comprehensive evaluation of these defenses is out of the scope of this paper. 
The design and evaluation of hybrid defenses are active research fields and we leave them for future work. 

\begin{figure*}[!tb]
    \begin{overpic}[width=\linewidth, height=0.5\linewidth]{figs/5x4.pdf}
        \put(0, 37){\includegraphics[width=0.095\linewidth]{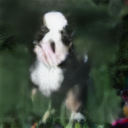}}
        \put(11, 37){\includegraphics[width=0.095\linewidth]{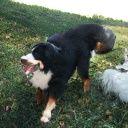}}
        \put(23, 37){\includegraphics[width=0.095\linewidth]{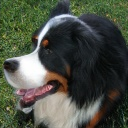}}
        \put(32.8, 37){\includegraphics[width=0.095\linewidth]{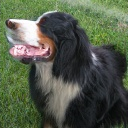}}
        \put(42.6, 37){\includegraphics[width=0.095\linewidth]{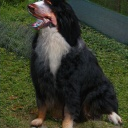}}
        \put(52.4, 37){\includegraphics[width=0.095\linewidth]{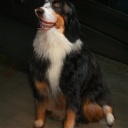}}
        \put(62.2, 37){\includegraphics[width=0.095\linewidth]{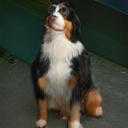}}
        \put(79.5, 37){\includegraphics[width=0.095\linewidth]{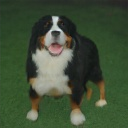}}
        \put(90.5, 37){\includegraphics[width=0.095\linewidth]{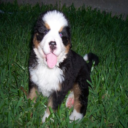}}

        \put(20.8, 41){\includegraphics[width=0.02\linewidth]{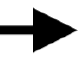}}
        \put(72.1, 41){\includegraphics[width=0.02\linewidth]{figs/lines/srarrow.png}}
        \put(74.5, 41.3){$\cdots$}
        \put(77.1, 41){\includegraphics[width=0.02\linewidth]{figs/lines/srarrow.png}}

        \put(0, 26.5){\includegraphics[width=0.095\linewidth]{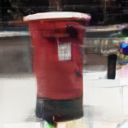}}
        \put(11, 26.5){\includegraphics[width=0.095\linewidth]{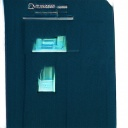}}
        \put(23, 26.5){\includegraphics[width=0.095\linewidth]{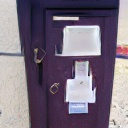}}
        \put(32.8, 26.5){\includegraphics[width=0.095\linewidth]{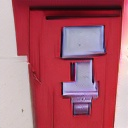}}
        \put(42.6, 26.5){\includegraphics[width=0.095\linewidth]{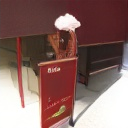}}
        \put(52.4, 26.5){\includegraphics[width=0.095\linewidth]{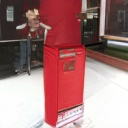}}
        \put(62.2, 26.5){\includegraphics[width=0.095\linewidth]{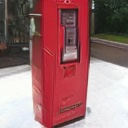}}
        \put(79.5, 26.5){\includegraphics[width=0.095\linewidth]{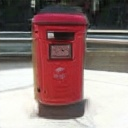}}
        \put(90.5, 26.5){\includegraphics[width=0.095\linewidth]{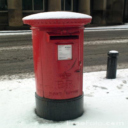}}

        \put(20.8, 30.5){\includegraphics[width=0.02\linewidth]{figs/lines/srarrow.png}}
        \put(72.1, 30.5){\includegraphics[width=0.02\linewidth]{figs/lines/srarrow.png}}
        \put(74.5, 30.8){$\cdots$}
        \put(77.1, 30.5){\includegraphics[width=0.02\linewidth]{figs/lines/srarrow.png}}

        \put(0, 16){\includegraphics[width=0.095\linewidth]{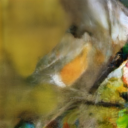}}
        \put(11, 16){\includegraphics[width=0.095\linewidth]{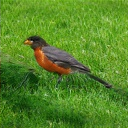}}
        \put(23, 16){\includegraphics[width=0.095\linewidth]{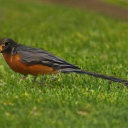}}
        \put(32.8, 16){\includegraphics[width=0.095\linewidth]{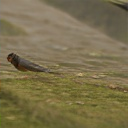}}
        \put(42.6, 16){\includegraphics[width=0.095\linewidth]{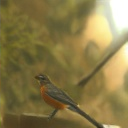}}
        \put(52.4, 16){\includegraphics[width=0.095\linewidth]{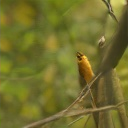}}
        \put(62.2, 16){\includegraphics[width=0.095\linewidth]{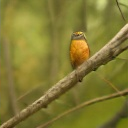}}
        \put(79.5, 16){\includegraphics[width=0.095\linewidth]{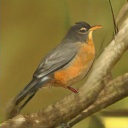}}
        \put(90.5, 16){\includegraphics[width=0.095\linewidth]{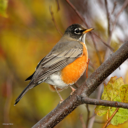}}

        \put(20.8, 20){\includegraphics[width=0.02\linewidth]{figs/lines/srarrow.png}}
        \put(72.1, 20){\includegraphics[width=0.02\linewidth]{figs/lines/srarrow.png}}
        \put(74.5, 20.3){$\cdots$}
        \put(77.1, 20){\includegraphics[width=0.02\linewidth]{figs/lines/srarrow.png}}

        \put(0, 5.5){\includegraphics[width=0.095\linewidth]{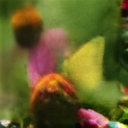}}
        \put(11, 5.5){\includegraphics[width=0.095\linewidth]{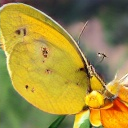}}
        \put(23, 5.5){\includegraphics[width=0.095\linewidth]{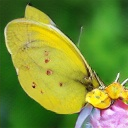}}
        \put(32.8, 5.5){\includegraphics[width=0.095\linewidth]{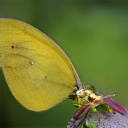}}
        \put(42.6, 5.5){\includegraphics[width=0.095\linewidth]{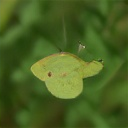}}
        \put(52.4, 5.5){\includegraphics[width=0.095\linewidth]{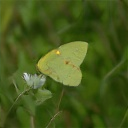}}
        \put(62.2, 5.5){\includegraphics[width=0.095\linewidth]{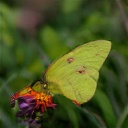}}
        \put(79.5, 5.5){\includegraphics[width=0.095\linewidth]{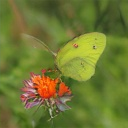}}
        \put(90.5, 5.5){\includegraphics[width=0.095\linewidth]{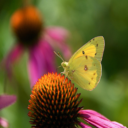}}

        \put(20.8, 9.5){\includegraphics[width=0.02\linewidth]{figs/lines/srarrow.png}}
        \put(72.1, 9.5){\includegraphics[width=0.02\linewidth]{figs/lines/srarrow.png}}
        \put(74.5, 9.8){$\cdots$}
        \put(77.1, 9.5){\includegraphics[width=0.02\linewidth]{figs/lines/srarrow.png}}

        \put(10.1, 5){\includegraphics[width=1.3pt, height=0.44\linewidth]{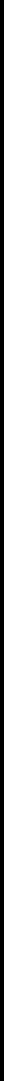}}
        \put(89.6, 5){\includegraphics[width=1.3pt, height=0.44\linewidth]{figs/lines/vline.png}}
        \put(.5, 48){\zoom[0.8]{Input $\hat{\X}_{m,i}$}}
        \put(11, 48){\zoom[0.8]{Initialization}}
        \put(40, 48){\zoom[0.8]{Search the Latent Space $\cdots$}}
        \put(79.5, 48){\zoom[0.8]{ROGS $\tilde{\X}_{m,i}$}}
        \put(92, 48){\zoom[0.8]{Raw $\X_{m,i}$}}
    \end{overpic}
    \vspace*{-35pt}
    \caption{
    Reconstruction attack at a semantic level using ROGS.
    Given lower-quality inputs $\hat{\X}_{m,i}$ from ROG, the optimization starts from a random latent vector corresponding to an image within the same class category as the ROG output. 
    The foreground and background will gradually change to match the lower-quality input during the latent-space search. 
    With some compromise at the pixel-level, the final results $\widetilde{\X}_{m,i}$ can better reveal the semantic visual information of the original images.   
    \label{fig:rogs_results}
    }
\end{figure*}

\section{Attack at a Semantic Level}\label{section:semnatic_attack}

In this section, we present a variant of the ROG attack by using a different postprocessing module. 
We propose to reconstruct a good-quality image at a \underline{\textbf{s}}emantic level (ROGS). 
Instead of focusing on the pixel-level error between the original image and the reconstructed counterpart, ROGS leverages a pretrained generative model $G$, such as BigGAN~\cite{brock2019large}, to synthesize realistic images while maintaining the semantics of the original image.  
The generative model $G$ maps a latent vector $\z$ to the image space, and the output has a similar distribution to the real images.       
We search the latent space of the generative model $G$ and restrict the similarity between the generated image and the original ROG lower-quality reconstruction. 
The goal of ROGS attack is to reconstruct an image sharing the same knowledge and semantics as in the original private image. 
Image reconstruction via the ROGS attack differs from the notion of property inference attack~\cite{melis2019exploiting} that target sensitive information. 
Sensitive information is generally difficult to define and corresponding privacy relies on personal preference~\cite{orekondy2017towards}. 
Compared to property inference, such as recovering gender, age, or race, ROGS may be viewed as a more general attack that diversifies the concept of compromising privacy. 
It has the capability to reveal privacy even if some sensitive information is not present in the corpus. 

From the technical perspective, we let ROGS invert an input to the GAN latent space, which is also known as the GAN inversion task in the literature~\cite{jahanian2019steerability, xia2021gan}. 
The optimization problem can be formulated as 
\begin{equation}\label{eq:rogs_problem}
    \z_{m,i}^\star = \argmin_{\z_{m,i}} \mathcal{L}(G(\mathbf{z}_{m,i}), \hat{\X}_{m,i}), 
\end{equation}
where $\mathcal{L}$ is a loss function quantifying the distance between two images, $\z$ denotes a latent vector, and $\hat{\X}$ is the original ROG output. 
The ROGS attack will take the latent vector $\z_{m,i}^\star$ as the input and output $G(\z_{m,i}^\star)$ as the attack result.

In our implementation, we use a combination of the $\ell_2$ distance and the LPIPS metric, namely, 
\begin{equation}
    \mathcal{L}(\X_1, \X_2) = \lambda_1 \, \cdot \| \X_1 - \X_2 \|_2 + \lambda_2 \, \cdot \mathrm{LPIPS} (\X_1, \X_2), 
\end{equation}
where $\lambda_1$ and $\lambda_2$ are the coefficients balancing the two terms. 
The loss function encourages the reconstructed image to preserve the structure with the $\ell_2$ distance and the perceptual styles of the original image with the $\mathrm{LPIPS}$ score. 
We choose the BigGAN model~\cite{brock2019large} as the generative model $G$, which is a class conditional GAN model trained on the ImageNet dataset. 
In particular, given a latent vector $\z_{m,i}$ and a class label $y_{m,i}$, the output of the model is constrained to be within the specific class. 
We start from a random initialization in the latent space and use the Adam optimizer to solve \eqref{eq:rogs_problem}. 

We use lower-quality images obtained in the ROG attack and show the attack results of ROGS in Figure~\ref{fig:rogs_results}.
It can be observed that the optimization process begins with a random initialization from the same image class, and gradually changes the foreground and background during the latent space search. 
In the first row, the raw image shows that a Bernese mountain dog is standing on the grass, facing the camera in the dark. 
The initialization chooses a dog of the same breed running on the grass in the daytime. 
During the optimization, the dog's posture and orientation are changed smoothly to match the lower-quality input. 
Meanwhile, the background is also gradually changed to the grass in the dark night. 
Although the reconstruction is not pixel-wise accurate, privacy leakage can still happen when rich semantics are revealed. 
Likewise, the details of the mailbox on the second row, including the shape, color, and layout, are adapted to match the counterpart in the input. 
The snow and the road in the background are reconstructed after the optimization. 
In the third row, the robin's size, coat color, and posture have been changed to a similar style as the original image after the latent space search, as well as the seasonal information. 
Similar changes can be observed for the sulphur butterfly picture on the fourth row, where the orientation and size of the butterfly are adjusted during the optimization. 

\begin{table*}[!tb]
    \centering
    \caption{
    Image pair and corresponding tags in the ROGS attack.
    \label{tab:rogs_results}
    }
    \begin{tabular}{m{0.12\textwidth}m{0.27\textwidth} m{0.12\textwidth}m{0.27\textwidth}}
        \toprule
        \hspace*{10pt} Image & \hspace*{40pt} Tags & \hspace*{10pt} Image & \hspace*{40pt} Tags \\ \midrule
        \includegraphics[width=0.11\textwidth]{figs/semantic/Bernese_mountain_dog/ori.png} & 
        {\footnotesize Black, Dog, White, Brown, Green, Snout, Grass, Tints and shades, Turquoise (Color), Habitat, Beige, Grey, Carnivore, Bernese mountain dog, Light, Limb, ...} &
        \includegraphics[width=0.11\textwidth]{figs/semantic/mailbox/ori.png} & 
        {\footnotesize Post box, Mailbox, Red, Daytime, Maroon, Infrastructure, Public space, White, Snow, Carmine, Line, Amber (Color), Morning, Circle, Material, ... } \\ \cmidrule{1-4}
        \includegraphics[width=0.11\textwidth]{figs/semantic/Bernese_mountain_dog/798.png} & 
        {\footnotesize Dog, Snout, Black, Brown, White, Bernese mountain dog, Green, Tints and shades, Habitat, Grass, Carnivore, Shoe, Nature, Turquoise (Color), ...} & 
        \includegraphics[width=0.11\textwidth]{figs/semantic/mailbox/595.png} & 
        {\footnotesize Post box, Red, Mailbox, Snow, Daytime, White, Maroon, Carmine, Public space, Composite material, Structure, Line, Product, Winter, ...} \\ \midrule
        \includegraphics[width=0.11\textwidth]{figs/semantic/robin/ori.png} & 
        {\footnotesize Bird, Brown, Beak, Orange (Color), Feather, Habitat, Daytime, Amber (Color), American robin, Morning, Branch, Twig, Ivory (Color), ...} & 
        \includegraphics[width=0.11\textwidth]{figs/semantic/sulphur_butterfly/ori.png} & 
        {\footnotesize Pollinator, Moths and butterflies, Flowering plant, Flower, Arthropod, Insect, Yellow, Butterfly, Magenta, Green, Purple coneflower, Spring (Season), ...} \\ \cmidrule{1-4}
        \includegraphics[width=0.11\textwidth]{figs/semantic/robin/924.png} & 
        {\footnotesize Bird, Beak, Brown, Feather, Habitat, Orange (Color), Amber (Color), Daytime, Ivory (Color), Ecoregion, Blond, Orange (Color), Peach (Color), ...} &
        \includegraphics[width=0.11\textwidth]{figs/semantic/sulphur_butterfly/983.png} & 
        {\footnotesize Pollinator, Arthropod, Moths and butterflies, Butterfly, Insect, Invertebrate, Yellow, Wing, Flowering plant, Green, Ecosystem, Computer wallpaper, ...} \\ \bottomrule
    \end{tabular}
\end{table*}

We now discuss how to quantify privacy leakage at the semantic level. 
We start by raising two concerns on current reconstruction evaluation metrics adopted in the literature based on the reconstruction results in Sections~\ref{section:rog}--\ref{section:current_defense}. 
First, we have observed that the attack success rate suggested by Wei et al.~\cite{wei2020framework} may not be a good indicator in some scenarios. 
For the reconstruction under existing defense schemes, privacy leakage should not be treated as a binary quantity. 
Merely using a threshold may not be a good indicator for a successful attack. 
The structural information or semantics can be revealed in the reconstructed image, although the results do not perfectly match the raw images.
There exists ambiguity in telling whether the attack is successful or not, let alone using an algorithm to automatically calculate the attack success rate. 
Second, the pixel-level based metrics, including MSE, PSNR, and SSIM, may fail to capture privacy leakage in some scenarios. 
For example, for the attack against signSGD in Figure~\ref{fig:attack_signsgd}, the average SSIM is $0.37$, which may not be able to fully capture the perceptual similarity due to the color jitter effect. 
A low PSNR value or a high MSE may give misleading conclusions on privacy leakage. 

To measure privacy leakage at a semantic level, we use a state-of-the-art multi-label classification network~\cite{ben2021multi} to tag the images.  
Specifically, the classification network has been trained on OpenImage (V6)~\cite{kuznetsova2020open}, which contains $9,\!600$ classes, including the object category, color, season, etc. 
Given two images $\X_1$ and $\X_2$, suppose their detected tags are denoted by set $A$ and $B$, respectively. 
The Jaccard similarity~\cite{jaccard1912distribution} between $A$ and $B$ is given as follows:
\begin{equation}
    J(A, B)=\frac{|A \cap B|}{|A \cup B|},
\end{equation}
where $|\cdot|$ denotes the cardinality of the set. 
The Jaccard similarity ranges from $0$ to $1$ by design. 

\begin{table}[!tb]
    \centering
    \caption{
    Comparisons between ROG and ROGS with different metrics.
    ROGS improves the semantic similarity score.
    \label{tab:rogs_metrics}
    }
    \begin{tabular}{llllll}
        \toprule
        Image & Attack & PSNR & SSIM & LPIPS & Jaccard \\ \midrule
        \multirow{2}{*}{Dog} & ROG & 22.1 dB & 0.897 & 0.332 & 0.400 \\
        & ROGS & 14.9 dB & 0.611 & 0.348 & 0.632 \\ \cmidrule{1-6}
        \multirow{2}{*}{Mailbox} & ROG & 21.3 dB & 0.946 & 0.206 & 0.433 \\
        & ROGS & 14.5 dB & 0.832 & 0.237 & 0.600 \\ \cmidrule{1-6}
        \multirow{2}{*}{Robin} & ROG & 20.8 dB & 0.836 & 0.299 & 0.250 \\
        & ROGS & 15.3 dB & 0.285 & 0.350 & 0.600 \\ \cmidrule{1-6}
        \multirow{2}{*}{Butterfly} & ROG & 21.3 dB & 0.900 & 0.287 & 0.312 \\
        & ROGS & 13.9 dB & 0.334 & 0.458 & 0.579 \\ \hdashline 
        \multicolumn{2}{l}{Random Noise$^1$} & 6.5 dB & 0.006 & 1.409 & 0.030 \\ 
        \multicolumn{2}{l}{Same Class Image$^2$} & 7.7 dB & 0.070 & 0.631 & 0.248 \\ \bottomrule
    \end{tabular}
    \begin{tablenotes}
    \small    
    \item \scriptsize $^1$ Average similarity between ROGS and random Gaussian noise over five repetitions. 
    \item \scriptsize $^2$ Average similarity between ROGS and images from the same ImageNet class over five repetitions. 
    \end{tablenotes}
\end{table}

The evaluation results of the same set of images in Figure~\ref{fig:rogs_results} are given in Table~\ref{tab:rogs_results} and Table~\ref{tab:rogs_metrics}. 
Table~\ref{tab:rogs_results} lists the pairs of the original image and the reconstructed one, along with the detected tags excluding some generic tags such as ``Photograph'', ``World'', and ``Color''. 
Table~\ref{tab:rogs_metrics} gives the reconstruction quality scores. 
It is observed that the tags of the dominating objects can be detected, including ``Bernese mountain dog'', ``Post box'', ``Bird'', and ``Butterfly''. 
The background information is also included in the tags, such as ``Grass'', ``Snow'', ``Twig'', and ``Purple coneflower''. 
The average PSNR, SSIM, and LPIPS scores compared to an image randomly selected from the same class are around $7.7$~dB,  $0.07$, $0.63$, respectively. 
PSNR values of ROGS results are around $14$~dB, which are worse than the ROG reconstruction and better than the random cases.
In addition, ROGS has slightly worse LPIPS values and worse SSIM compared to ROG. 
In other words, ROGS does not directly improve the image similarity at the pixel level as compared to ROG, but consistently outperforms ROG in terms of Jaccard similarity scores, which is more oriented toward recognition at the semantic level. 
We now compare the difference between the SSIM value and the Jaccard similarity. 
For the first pair of Bernese mountain dogs, SSIM and Jaccard values are above $0.6$ and tend to agree with each other.  
For the second pair of post boxes, the SSIM gives a much higher value, $0.832$, compared to Jaccard similarity, $0.6$. 
A large discrepancy between SSIM and Jaccard can be found in the third and fourth pairs. 
For the third pair of robins, the SSIM is $0.285$, which is relatively low. 
In the meantime, the semantics have been successfully reconstructed, which is also reflected in a Jaccard similarity of $0.6$. 
A similar pattern can be observed in the fourth pair. 
The reconstruction has revealed important information about butterflies resting on the flower. 
Such a privacy loss may be ignored when merely checking the SSIM value of $0.334$.

\begin{figure}[!tb]
    \begin{overpic}[width=\linewidth, height=0.5\linewidth]{figs/5x4.pdf}
        \put(20, 30){\includegraphics[width=0.19\linewidth]{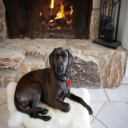}}
        \put(39.5, 30){\includegraphics[width=0.19\linewidth]{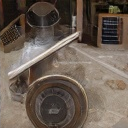}}
        \put(61, 30){\includegraphics[width=0.19\linewidth]{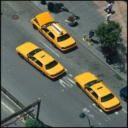}}
        \put(80.5, 30){\includegraphics[width=0.19\linewidth]{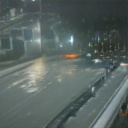}}
    
        \put(20, 7){\includegraphics[width=0.19\linewidth]{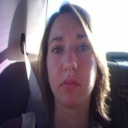}}
        \put(39.5, 7){\includegraphics[width=0.19\linewidth]{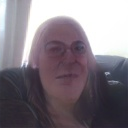}}
        \put(61, 7){\includegraphics[width=0.19\linewidth]{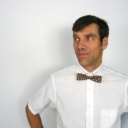}}
        \put(80.5, 7){\includegraphics[width=0.19\linewidth]{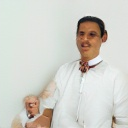}}
    
        \put(0,40){\intab{Complicated}}
        \put(0,36){\intab{Scenarios}}
        \put(0,18){\intab{Human}}
        \put(0,14){\intab{Faces}}  
    \end{overpic}
    \vspace*{-30pt}
    \caption{
    Image pairs of the raw data and the reconstruction that deviate at the semantic level.
    On the first row, ROGS fails to reconstruct images containing complicated scenarios. 
    For the second row involving human faces, the reconstruction fails to reveal their identities. 
    \label{fig:negative_examples}
    }
\end{figure}
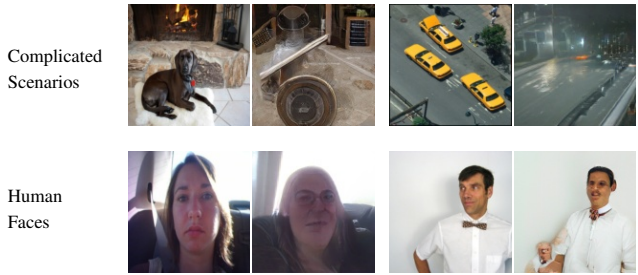

\highlight{Limitations. }
We would like to clarify that the aforementioned Jaccard similarity is not intended to replace the current metrics for image privacy loss measurement, but rather to serve as a supplement or as a motivation to help us rethink the privacy leakage problem. 
In the meantime, there still exist some limitations in the reconstruction and evaluation at the semantic level.
We give some negative reconstruction examples in Figure~\ref{fig:negative_examples}. 
When the raw images contain a complicated scenario, such as including multiple dominating objects, the attack algorithm may fail to converge. 
This can be observed in the first row of Figure~\ref{fig:negative_examples}.
Another example is when the raw image involves human faces, the reconstruction may not be able to reveal the person's identity.
This may not be considered as a severe privacy leakage in some applications, such as facial data analysis. 
Furthermore, not all of the semantics can be included in the tags, such as the action, orientation, and body size. 
In addition, the multi-label classification neural network may not always give the correct tags. 
For example, a tag ``Shoe'' is detected in the second row of Table~\ref{tab:rogs_results}, which can be considered as a misclassification. 
The importance of different tags and their correlations may also need to be considered to improve the similarity score design. 
A more comprehensive study is out of the scope of this work. 
However, we hope these results can inspire researchers to revisit the privacy leakage issues in federated learning.

\section{Discussion and Future Work}\label{section:discussion}

The concept of federated learning was proposed to preserve sensitive client data for multi-party machine learning. 
In this work, we have shown that it is possible to reconstruct client data with high quality from noisy gradients in a more realistic federated learning setting compared to existing attacks. 
Contrary to prior empirical work, we have demonstrated that gradient compression and perturbation cannot be treated as effective privacy protection strategies. 
Our attack scheme has also successfully reconstructed private information under existing defenses. 
Based on these observations, we conjecture that for federated learning algorithms that do not provide privacy guarantees for data attributes, an adversary can always find a certain attack scheme to disclose the privacy in the raw data, as long as the mutual information between gradient and raw data is not close to zero. 

In the future, it will be interesting to investigate the trade-off between utility and privacy for hybrid defenses.
As it has been observed in~\cite{huang2021evaluating}, a combination of multiple defenses can provide better privacy protection. 
Besides, differentially private training with secure aggregation~\cite{agarwal2018cpsgd, kairouz2021distributed} may also have potential to defend against the proposed ROG attack.
In parallel to the existing differential privacy paradigm, a more general formulation of the privacy concept will be worth exploring. 
We also believe that the study of privacy leakage beyond the image classification task will provide more insights to the research community. 
\bibliographystyle{plain}
\bibliography{myref}

\begin{thebibliography}{10}

\bibitem{abadi2016deep}
Martin Abadi, Andy Chu, Ian Goodfellow, H~Brendan McMahan, Ilya Mironov, Kunal
  Talwar, and Li~Zhang.
\newblock Deep learning with differential privacy.
\newblock In {\em ACM SIGSAC Conference on Computer and Communications
  Security}, 2016.

\bibitem{agarwal2018cpsgd}
Naman Agarwal, Ananda~Theertha Suresh, Felix Xinnan~X Yu, Sanjiv Kumar, and
  Brendan McMahan.
\newblock cp{SGD}: Communication-efficient and differentially-private
  distributed {SGD}.
\newblock {\em Advances in Neural Information Processing Systems}, 31, 2018.

\bibitem{aji2017sparse}
Alham~Fikri {Aji} and Kenneth {Heafield}.
\newblock Sparse communication for distributed gradient descent.
\newblock In {\em Conference on Empirical Methods in Natural Language
  Processing}, pages 440--445, 2017.

\bibitem{alistarh2017qsgd}
Dan {Alistarh}, Demjan {Grubic}, Jerry {Li}, Ryota {Tomioka}, and Milan
  {Vojnovic}.
\newblock Q{SGD}: Communication-efficient {SGD} via gradient quantization and
  encoding.
\newblock In {\em Advances in Neural Information Processing Systems}, pages
  1709--1720, 2017.

\bibitem{baldi2012autoencoders}
Pierre Baldi.
\newblock Autoencoders, unsupervised learning, and deep architectures.
\newblock In {\em Proceedings of ICML Workshop on Unsupervised and Transfer
  Learning}, pages 37--49. JMLR Workshop and Conference Proceedings, 2012.

\bibitem{balunovic2021bayesian}
Mislav Balunovi{\'c}, Dimitar~I Dimitrov, Robin Staab, and Martin Vechev.
\newblock Bayesian framework for gradient leakage.
\newblock In {\em International Conference on Learning Representations}, 2022.

\bibitem{ben2021multi}
Emanuel Ben-Baruch, Tal Ridnik, Itamar Friedman, Avi Ben-Cohen, Nadav Zamir,
  Asaf Noy, and Lihi Zelnik-Manor.
\newblock Multi-label classification with partial annotations using class-aware
  selective loss.
\newblock {\em arXiv preprint arXiv:2110.10955}, 2021.

\bibitem{bernstein2018signsgd}
Jeremy Bernstein, Yu-Xiang Wang, Kamyar Azizzadenesheli, and Animashree
  Anandkumar.
\newblock {S}ign{SGD}: Compressed optimisation for non-convex problems.
\newblock In {\em International Conference on Machine Learning}, pages
  560--569, 2018.

\bibitem{bonawitz2016practical}
Keith Bonawitz, Vladimir Ivanov, Ben Kreuter, Antonio Marcedone, H~Brendan
  McMahan, Sarvar Patel, Daniel Ramage, Aaron Segal, and Karn Seth.
\newblock Practical secure aggregation for federated learning on user-held
  data.
\newblock {\em NeurIPS 2016 workshop on Private Multi-Party Machine Learning},
  2016.

\bibitem{boutet2021mixnn}
Antoine Boutet, Thomas Lebrun, Jan Aalmoes, and Adrien Baud.
\newblock Mix{NN}: Protection of federated learning against inference attacks
  by mixing neural network layers.
\newblock {\em arXiv preprint arXiv:2109.12550}, 2021.

\bibitem{brock2019large}
Andrew Brock, Jeff Donahue, and Karen Simonyan.
\newblock Large scale {GAN} training for high fidelity natural image synthesis.
\newblock In {\em International Conference on Learning Representations}, 2019.

\bibitem{carlini2021private}
Nicholas Carlini, Samuel Deng, Sanjam Garg, Somesh Jha, Saeed Mahloujifar,
  Mohammad Mahmoody, Abhradeep Thakurta, and Florian Tram{\`e}r.
\newblock Is private learning possible with instance encoding?
\newblock In {\em 2021 IEEE Symposium on Security and Privacy (SP)}, pages
  410--427. IEEE, 2021.

\bibitem{cheng2021separation}
Pau-Chen Cheng, Kevin Eykholt, Zhongshu Gu, Hani Jamjoom, KR~Jayaram,
  Enriquillo Valdez, and Ashish Verma.
\newblock Separation of powers in federated learning.
\newblock {\em arXiv preprint arXiv:2105.09400}, 2021.

\bibitem{deng2009imagenet}
Jia Deng, Wei Dong, Richard Socher, Li-Jia Li, Kai Li, and Li~Fei-Fei.
\newblock Image{N}et: A large-scale hierarchical image database.
\newblock In {\em IEEE Conference on Computer Vision and Pattern Recognition},
  pages 248--255. IEEE, 2009.

\bibitem{deng2021tag}
Jieren Deng, Yijue Wang, Ji~Li, Chenghong Wang, Chao Shang, Hang Liu,
  Sanguthevar Rajasekaran, and Caiwen Ding.
\newblock Tag: Gradient attack on transformer-based language models.
\newblock In {\em Findings of the Association for Computational Linguistics},
  pages 3600--3610, 2021.

\bibitem{fowl2021robbing}
Liam Fowl, Jonas Geiping, Wojtek Czaja, Micah Goldblum, and Tom Goldstein.
\newblock Robbing the fed: Directly obtaining private data in federated
  learning with modified models.
\newblock {\em arXiv preprint arXiv:2110.13057}, 2021.

\bibitem{geiping2020inverting}
Jonas {Geiping}, Hartmut {Bauermeister}, Hannah {Dröge}, and Michael
  {Moeller}.
\newblock Inverting gradients - how easy is it to break privacy in federated
  learning?
\newblock In {\em Advances in Neural Information Processing Systems}, 2020.

\bibitem{gilad2016cryptonets}
Ran Gilad-Bachrach, Nathan Dowlin, Kim Laine, Kristin Lauter, Michael Naehrig,
  and John Wernsing.
\newblock Crypto{N}ets: Applying neural networks to encrypted data with high
  throughput and accuracy.
\newblock In {\em International Conference on Machine Learning}, 2016.

\bibitem{girgis2021shuffled}
Antonious Girgis, Deepesh Data, Suhas Diggavi, Peter Kairouz, and
  Ananda~Theertha Suresh.
\newblock Shuffled model of differential privacy in federated learning.
\newblock In {\em International Conference on Artificial Intelligence and
  Statistics}, 2021.

\bibitem{gonzalez2014digital}
Rafael~C. {Gonzalez} and Richard~E. {Woods}.
\newblock {\em Digital {I}mage {P}rocessing, 3rd {E}dition}.
\newblock 2014.

\bibitem{good1976application}
Irving~J Good.
\newblock On the application of symmetric {D}irichlet distributions and their
  mixtures to contingency tables.
\newblock {\em The Annals of Statistics}, 4(6):1159--1189, 1976.

\bibitem{gu2021federated}
Hanlin Gu, Lixin Fan, Bowen Li, Yan Kang, Yuan Yao, and Qiang Yang.
\newblock Federated deep learning with {B}ayesian privacy.
\newblock {\em arXiv preprint arXiv:2109.13012}, 2021.

\bibitem{he2016deep}
Kaiming He, Xiangyu Zhang, Shaoqing Ren, and Jian Sun.
\newblock Deep residual learning for image recognition.
\newblock In {\em IEEE conference on computer vision and pattern recognition},
  pages 770--778, 2016.

\bibitem{hsieh2020non}
Kevin Hsieh, Amar Phanishayee, Onur Mutlu, and Phillip Gibbons.
\newblock The non-iid data quagmire of decentralized machine learning.
\newblock In {\em International Conference on Machine Learning}, pages
  4387--4398, 2020.

\bibitem{hsu2019measuring}
Tzu-Ming~Harry Hsu, Hang Qi, and Matthew Brown.
\newblock Measuring the effects of non-identical data distribution for
  federated visual classification.
\newblock {\em arXiv preprint arXiv:1909.06335}, 2019.

\bibitem{huang2021evaluating}
Yangsibo Huang, Samyak Gupta, Zhao Song, Kai Li, and Sanjeev Arora.
\newblock Evaluating gradient inversion attacks and defenses in federated
  learning.
\newblock {\em Advances in Neural Information Processing Systems}, 34, 2021.

\bibitem{huang2020instahide}
Yangsibo Huang, Zhao Song, Kai Li, and Sanjeev Arora.
\newblock Instahide: Instance-hiding schemes for private distributed learning.
\newblock In {\em International conference on machine learning}, pages
  4507--4518. PMLR, 2020.

\bibitem{isola2017image}
Phillip Isola, Jun-Yan Zhu, Tinghui Zhou, and Alexei~A Efros.
\newblock Image-to-image translation with conditional adversarial networks.
\newblock In {\em Proceedings of the IEEE conference on computer vision and
  pattern recognition}, pages 1125--1134, 2017.

\bibitem{jaccard1912distribution}
Paul Jaccard.
\newblock The distribution of the flora in the alpine zone. 1.
\newblock {\em New Phytologist}, 11(2):37--50, 1912.

\bibitem{jahanian2019steerability}
Ali Jahanian, Lucy Chai, and Phillip Isola.
\newblock On the ``steerability" of generative adversarial networks.
\newblock In {\em International Conference on Learning Representations}, 2019.

\bibitem{jeon2021gradient}
Jinwoo Jeon, Kangwook Lee, Sewoong Oh, Jungseul Ok, et~al.
\newblock Gradient inversion with generative image prior.
\newblock {\em Advances in Neural Information Processing Systems},
  34:29898--29908, 2021.

\bibitem{jere2020taxonomy}
Malhar~S Jere, Tyler Farnan, and Farinaz Koushanfar.
\newblock A taxonomy of attacks on federated learning.
\newblock {\em IEEE Security \& Privacy}, 19(2):20--28, 2020.

\bibitem{kairouz2021distributed}
Peter Kairouz, Ziyu Liu, and Thomas Steinke.
\newblock The distributed discrete gaussian mechanism for federated learning
  with secure aggregation.
\newblock In {\em International Conference on Machine Learning}, pages
  5201--5212. PMLR, 2021.

\bibitem{kairouz2021advances}
Peter Kairouz, H~Brendan McMahan, Brendan Avent, Aur{\'e}lien Bellet, Mehdi
  Bennis, Arjun~Nitin Bhagoji, Keith Bonawitz, Zachary Charles, Graham Cormode,
  Rachel Cummings, et~al.
\newblock Advances and open problems in federated learning.
\newblock {\em Foundations and Trends in Machine Learning}, 14(1), 2021.

\bibitem{kerkouche2021compression}
Raouf {Kerkouche}, Gergely {Ács}, Claude {Castelluccia}, and Pierre
  {Genevès}.
\newblock Compression boosts differentially private federated learning.
\newblock In {\em European Symposium on Security and Privacy}, pages 1--15,
  2021.

\bibitem{kingma2015adam}
Diederik~P Kingma and Jimmy Ba.
\newblock Adam: A method for stochastic optimization.
\newblock 2015.

\bibitem{krizhevsky2009learning}
Alex {Krizhevsky}.
\newblock Learning multiple layers of features from tiny images.
\newblock {\em Master thesis, Dept. of Comput. Sci., Univ. of Toronto, Toronto,
  Canada}, 2009.

\bibitem{kuznetsova2020open}
Alina Kuznetsova, Hassan Rom, Neil Alldrin, Jasper Uijlings, Ivan Krasin, Jordi
  Pont-Tuset, Shahab Kamali, Stefan Popov, Matteo Malloci, Alexander
  Kolesnikov, et~al.
\newblock The open images dataset v4.
\newblock {\em International Journal of Computer Vision}, 128(7):1956--1981,
  2020.

\bibitem{lam2021gradient}
Maximilian Lam, Gu-Yeon Wei, David Brooks, Vijay~Janapa Reddi, and Michael
  Mitzenmacher.
\newblock Gradient disaggregation: Breaking privacy in federated learning by
  reconstructing the user participant matrix.
\newblock In {\em International Conference on Machine Learning}, pages
  5959--5968. PMLR, 2021.

\bibitem{li2020fedprox}
Tian {Li}, Anit~Kumar {Sahu}, Manzil {Zaheer}, Maziar {Sanjabi}, Ameet
  {Talwalkar}, and Virginia {Smith}.
\newblock Federated optimization in heterogeneous networks.
\newblock {\em Proceedings of Machine Learning and Systems}, 2:429--450, 2020.

\bibitem{liu2021quantitative}
Yong Liu, Xinghua Zhu, Jianzong Wang, and Jing Xiao.
\newblock A quantitative metric for privacy leakage in federated learning.
\newblock In {\em IEEE International Conference on Acoustics, Speech and Signal
  Processing}, 2021.

\bibitem{lyu2020threats}
Lingjuan Lyu, Han Yu, and Qiang Yang.
\newblock Threats to federated learning: A survey.
\newblock {\em arXiv preprint arXiv:2003.02133}, 2020.

\bibitem{mcmahan2017communication}
H.~Brendan {McMahan}, Eider {Moore}, Daniel {Ramage}, Seth {Hampson}, and
  Blaise~Aguera y~{Arcas}.
\newblock Communication-efficient learning of deep networks from decentralized
  data.
\newblock In {\em International Conference on Artificial Intelligence and
  Statistics}, 2017.

\bibitem{melis2019exploiting}
Luca Melis, Congzheng Song, Emiliano De~Cristofaro, and Vitaly Shmatikov.
\newblock Exploiting unintended feature leakage in collaborative learning.
\newblock In {\em 2019 IEEE symposium on security and privacy (SP)}, pages
  691--706. IEEE, 2019.

\bibitem{mo2021quantifying}
Fan Mo, Anastasia Borovykh, Mohammad Malekzadeh, Hamed Haddadi, and Soteris
  Demetriou.
\newblock Quantifying and localizing private information leakage from neural
  network gradients.
\newblock {\em arXiv preprint arXiv:2105.13929}, 2021.

\bibitem{nazeri2019edgeconnect}
Kamyar Nazeri, Eric Ng, Tony Joseph, Faisal~Z Qureshi, and Mehran Ebrahimi.
\newblock Edgeconnect: Generative image inpainting with adversarial edge
  learning.
\newblock {\em arXiv preprint arXiv:1901.00212}, 2019.

\bibitem{nguyen2021federated}
Dinh~C Nguyen, Ming Ding, Quoc-Viet Pham, Pubudu~N Pathirana, Long~Bao Le,
  Aruna Seneviratne, Jun Li, Dusit Niyato, and H~Vincent Poor.
\newblock Federated learning meets blockchain in edge computing: Opportunities
  and challenges.
\newblock {\em IEEE Internet of Things Journal}, 2021.

\bibitem{orekondy2017towards}
Tribhuvanesh Orekondy, Bernt Schiele, and Mario Fritz.
\newblock Towards a visual privacy advisor: Understanding and predicting
  privacy risks in images.
\newblock In {\em Proceedings of the IEEE international conference on computer
  vision}, pages 3686--3695, 2017.

\bibitem{pan2020theory}
Xudong Pan, Mi~Zhang, Yifan Yan, Jiaming Zhu, and Min Yang.
\newblock Theory-oriented deep leakage from gradients via linear equation
  solver.
\newblock {\em arXiv preprint arXiv:2010.13356}, 2020.

\bibitem{qian2020minimal}
Jia Qian, Hiba Nassar, and Lars~Kai Hansen.
\newblock Minimal conditions analysis of gradient-based reconstruction in
  federated learning.
\newblock {\em arXiv preprint arXiv:2010.15718}, 2020.

\bibitem{reisizadeh2020fedpaq}
Amirhossein {Reisizadeh}, Aryan {Mokhtari}, Hamed {Hassani}, Ali {Jadbabaie},
  and Ramtin {Pedarsani}.
\newblock Fed{PAQ}: A communication-efficient federated learning method with
  periodic averaging and quantization.
\newblock In {\em International Conference on Artificial Intelligence and
  Statistics}, 2020.

\bibitem{rigaki2020survey}
Maria Rigaki and Sebastian Garcia.
\newblock A survey of privacy attacks in machine learning.
\newblock {\em arXiv preprint arXiv:2007.07646}, 2020.

\bibitem{ronneberger2015u}
Olaf Ronneberger, Philipp Fischer, and Thomas Brox.
\newblock U-net: Convolutional networks for biomedical image segmentation.
\newblock In {\em International Conference on Medical image computing and
  computer-assisted intervention}, pages 234--241. Springer, 2015.

\bibitem{scheliga2022precode}
Daniel Scheliga, Patrick M{\"a}der, and Marco Seeland.
\newblock {PRECODE} -- a generic model extension to prevent deep gradient
  leakage.
\newblock In {\em IEEE/CVF Winter Conference on Applications of Computer
  Vision}, pages 1849--1858, 2022.

\bibitem{simonyan2015very}
Karen Simonyan and Andrew Zisserman.
\newblock Very deep convolutional networks for large-scale image recognition.
\newblock 2015.

\bibitem{sun2021soteria}
Jingwei Sun, Ang Li, Binghui Wang, Huanrui Yang, Hai Li, and Yiran Chen.
\newblock Soteria: Provable defense against privacy leakage in federated
  learning from representation perspective.
\newblock In {\em IEEE/CVF Conference on Computer Vision and Pattern
  Recognition}, pages 9311--9319, 2021.

\bibitem{wainakh2021label}
Aidmar Wainakh, Till M{\"u}{\ss}ig, Tim Grube, and Max M{\"u}hlh{\"a}user.
\newblock Label leakage from gradients in distributed machine learning.
\newblock In {\em Annual Consumer Communications \& Networking Conference},
  2021.

\bibitem{wang2021a}
Jianyu {Wang}, Zachary {Charles}, Zheng {Xu}, Gauri {Joshi}, H.~Brendan
  {McMahan}, Blaise~Aguera y~{Arcas}, et~al.
\newblock A field guide to federated optimization.
\newblock {\em arXiv preprint arXiv:2107.06917}, 2021.

\bibitem{wang2004image}
Zhou Wang, Alan~C Bovik, Hamid~R Sheikh, and Eero~P Simoncelli.
\newblock Image quality assessment: from error visibility to structural
  similarity.
\newblock {\em IEEE transactions on image processing}, 13(4):600--612, 2004.

\bibitem{wei2021user}
Kang {Wei}, Jun {Li}, Ming {Ding}, Chuan {Ma}, Hang {Su}, Bo~{Zhang}, and
  H.~Vincent {Poor}.
\newblock User-level privacy-preserving federated learning: Analysis and
  performance optimization.
\newblock {\em IEEE Transactions on Mobile Computing}, 2021.

\bibitem{wei2020framework}
Wenqi Wei, Ling Liu, Margaret Loper, Ka-Ho Chow, Mehmet~Emre Gursoy, Stacey
  Truex, and Yanzhao Wu.
\newblock A framework for evaluating client privacy leakages in federated
  learning.
\newblock In {\em Computer Security – ESORICS 2020: 25th European Symposium
  on Research in Computer Security, ESORICS 2020, Guildford, UK, September
  14–18, 2020, Proceedings, Part I}, page 545–566, Berlin, Heidelberg,
  2020. Springer-Verlag.

\bibitem{wei2021gradient}
Wenqi Wei, Ling Liu, Yanzhao Wut, Gong Su, and Arun Iyengar.
\newblock Gradient-leakage resilient federated learning.
\newblock In {\em International Conference on Distributed Computing Systems},
  2021.

\bibitem{wen2022fishing}
Yuxin Wen, Jonas Geiping, Liam Fowl, Micah Goldblum, and Tom Goldstein.
\newblock Fishing for user data in large-batch federated learning via gradient
  magnification.
\newblock In {\em International Conference on Machine Learning}, pages
  23668--23684. PMLR, 2022.

\bibitem{xia2021gan}
Weihao Xia, Yulun Zhang, Yujiu Yang, Jing-Hao Xue, Bolei Zhou, and Ming-Hsuan
  Yang.
\newblock {GAN} inversion: A survey.
\newblock {\em arXiv preprint arXiv:2101.05278}, 2021.

\bibitem{yin2021see}
Hongxu Yin, Arun Mallya, Arash Vahdat, Jose~M Alvarez, Jan Kautz, and Pavlo
  Molchanov.
\newblock See through gradients: Image batch recovery via gradinversion.
\newblock In {\em IEEE/CVF Conference on Computer Vision and Pattern
  Recognition}, pages 16337--16346, 2021.

\bibitem{yu2021do}
Da~{Yu}, Huishuai {Zhang}, Wei {Chen}, and Tie-Yan {Liu}.
\newblock Do not let privacy overbill utility: Gradient embedding perturbation
  for private learning.
\newblock In {\em International Conference on Learning Representations}, 2021.

\bibitem{zhang2020batchcrypt}
Chengliang Zhang, Suyi Li, Junzhe Xia, Wei Wang, Feng Yan, and Yang Liu.
\newblock Batch{C}rypt: Efficient homomorphic encryption for cross-silo
  federated learning.
\newblock In {\em {USENIX} Annual Technical Conference}, pages 493--506, 2020.

\bibitem{zhang2021designing}
Kai Zhang, Jingyun Liang, Luc Van~Gool, and Radu Timofte.
\newblock Designing a practical degradation model for deep blind image
  super-resolution.
\newblock In {\em IEEE/CVF International Conference on Computer Vision}, pages
  4791--4800, 2021.

\bibitem{zhang2018unreasonable}
Richard Zhang, Phillip Isola, Alexei~A Efros, Eli Shechtman, and Oliver Wang.
\newblock The unreasonable effectiveness of deep features as a perceptual
  metric.
\newblock In {\em IEEE {C}onference on {C}omputer {V}ision and {P}attern
  {R}ecognition}, pages 586--595, 2018.

\bibitem{zhang2022survey}
Rui Zhang, Song Guo, Junxiao Wang, Xin Xie, and Dacheng Tao.
\newblock A survey on gradient inversion: Attacks, defenses and future
  directions.
\newblock {\em arXiv preprint arXiv:2206.07284}, 2022.

\bibitem{zhang2020secret}
Yuheng Zhang, Ruoxi Jia, Hengzhi Pei, Wenxiao Wang, Bo~Li, and Dawn Song.
\newblock The secret revealer: Generative model-inversion attacks against deep
  neural networks.
\newblock In {\em IEEE/CVF Conference on Computer Vision and Pattern
  Recognition}, pages 253--261, 2020.

\bibitem{zhao2020idlg}
Bo~{Zhao}, Konda~Reddy {Mopuri}, and Hakan {Bilen}.
\newblock i{DLG}: Improved deep leakage from gradients.
\newblock {\em arXiv preprint arXiv:2001.02610}, 2020.

\bibitem{zhu2019deep}
Ligeng Zhu, Zhijian Liu, and Song Han.
\newblock Deep leakage from gradients.
\newblock In {\em Advances in Neural Information Processing Systems}, 2019.

\end{thebibliography}

\appendices

\section{Implementation Details}\label{section:implementation}
We provide more details of the postprocessing neural networks as follows. 
We adopt the U-Net architecture~\cite{ronneberger2015u} as the GAN generator, which is composed of an encoder that downsamples the images and a decoder that upsamples the feature maps back to the original size.
This architecture has been widely adopted in image-to-image translation~\cite{isola2017image} and image inpainting~\cite{nazeri2019edgeconnect}.  
In the generator, we take the form of convolution-InstanceNorm-ReLu for the backbone module, and downsample the input image four times.  
We add skip connections between the encoder and the decoder to circumvent severe information loss caused by the downsampling.
The discriminator architecture is based on PatchGAN~\cite{nazeri2019edgeconnect}. 

To synthesize the dataset, we randomly apply one of the downsampling approaches, including nearest neighbor, bilinear scaling, and bicubic scaling,  and add Gaussian noise to the training examples in the ImageNet dataset.
In particular, we randomly choose the order of downsampling, noise injection, and a Gaussian blur kernel to degrade the original images. 
We use a linear combination of the $\ell_1$ loss and the adversarial loss as the cost function. 
The Adam optimizer is adopted for the GAN training, and the learning rate is set to $1\times 10^{-5}$ for both the generator and the discriminator. 
For ROGS, we set $\lambda_1 = \lambda_2 = 1$ and optimize for $1000$ iterations. 
The pretrained models and implementation of the attack are available at \url{https://anonymous.4open.science/r/rog-DC5E}.

\section{Image Quality Metrics}\label{app:IQA}
We review the image quality metrics used in this work, including PSNR, SSIM, and LPIPS.
Given two images $\X, \Y \in \mathbb{R}^{h\times w}$, the mean squared error (MSE) is defined as:
\begin{equation}
    \mathrm{MSE}(\X,\Y) = \frac{1}{h\,w} \sum_{i=1}^h\sum_{j=1}^w (X_{i,j} - Y_{i,j})^2,
\end{equation}
where $X_{i,j}$ denotes the $(i,j)$th entry of the matrix $\X$. 
The PSNR in dB is given by:
\begin{equation}
    \mathrm{P S N R}(\X, \Y) = 10 \cdot \log _{10}\left(\frac{\mathrm{M A X}_{I}^{2}}{\mathrm{MSE}}\right),
\end{equation}
where $\mathrm{MAX}_I$ is the maximum possible pixel value of the image.
By definition, PSNR is a pixel-wise similarity metric.
SSIM better approximates the perception model. 
Given two images $\X, \Y$, suppose $\mu_X$ and $\mu_{Y}$ are the average of $\X$ and $\Y$, $\sigma_{X}^2$ and $\sigma_{Y}^2$ are the variance of the two images, and $\sigma_{XY}$ is the covariance.
The SSIM is given by:
\begin{equation}
    \operatorname{SSIM}(\X, \Y)=\frac{\left(2 \mu_{X} \mu_{Y}+c_{1}\right)\left(2 \sigma_{X Y} +c_{2}\right)}{\left(\mu_{X}^{2}+\mu_{Y}^{2}+c_{1}\right)\left(\sigma_{X}^{2}+\sigma_{Y}^{2}+c_{2}\right)}, 
\end{equation}
where $c_1$ and $c_2$ are two variables to stabilize the division. 
By design, $\operatorname{SSIM} \in [0,1]$, and a higher value indicates a higher similarity. 

Compared to the traditional image quality metrics, LPIPS was proposed based on the activations of convolutional neural networks.
Consider the $l$th layer of a neural network with unit-normalized feature $\boldsymbol{A}^l \in \mathbb{R}^{H_l \times W_l \times C_l}$, where $H_l$, $W_l$, and $C_l$ denote its height, width, and channels, respectively. 
With predefined coefficients $\mathbf{c}_l \in \mathbb{R}^{C_l}$, the LPIPS value can be calculated as 
\begin{equation}
    \mathrm{LPIPS}(\X, \Y) = \sum_l \frac{1}{H_l W_l} \sum_{h, w}  \| \mathbf{c}_l \odot (\boldsymbol{A}^{l}_{h,w} - \boldsymbol{B}^{l}_{h,w}\|^2_2, 
\end{equation}
where $\boldsymbol{A}^{l}_{h,w}$ and $\boldsymbol{B}^{l}_{h,w}$ denote the feature maps of two images.

\section{More Reconstruction Examples}\label{section:more_example}

\begin{figure}[!tb]
    \subcaptionbox{\label{subfig:ori_fullbatch2}}[0.495\linewidth]{
    \begin{overpic}[width=\linewidth, height=\linewidth]{figs/4x4.pdf}
    \put(0, 75){\includegraphics[width=0.25\linewidth]{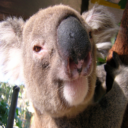}}
    \put(25, 75){\includegraphics[width=0.25\linewidth]{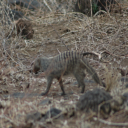}}
    \put(50, 75){\includegraphics[width=0.25\linewidth]{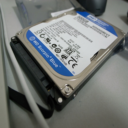}}
    \put(75, 75){\includegraphics[width=0.25\linewidth]{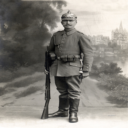}}

    \put(0, 50){\includegraphics[width=0.25\linewidth]{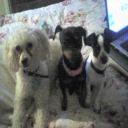}}
    \put(25, 50){\includegraphics[width=0.25\linewidth]{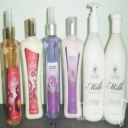}}
    \put(50, 50){\includegraphics[width=0.25\linewidth]{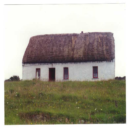}}
    \put(75, 50){\includegraphics[width=0.25\linewidth]{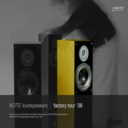}}

    \put(0, 25){\includegraphics[width=0.25\linewidth]{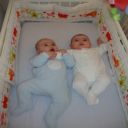}}
    \put(25, 25){\includegraphics[width=0.25\linewidth]{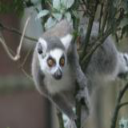}}
    \put(50, 25){\includegraphics[width=0.25\linewidth]{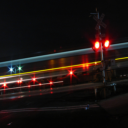}}
    \put(75, 25){\includegraphics[width=0.25\linewidth]{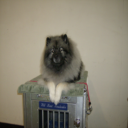}}
    
    \put(0, 0){\includegraphics[width=0.25\linewidth]{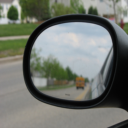}}
    \put(25, 0){\includegraphics[width=0.25\linewidth]{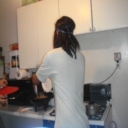}}
    \put(50, 0){\includegraphics[width=0.25\linewidth]{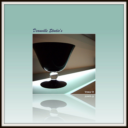}}
    \put(75, 0){\includegraphics[width=0.25\linewidth]{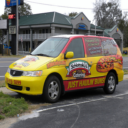}}
    \end{overpic}
    }
    \subcaptionbox{\label{subfig:recon_fullbatch_quant}}[0.495\linewidth]{
        \begin{overpic}[width=\linewidth, height=\linewidth]{figs/4x4.pdf}
        \put(0, 75){\includegraphics[width=0.25\linewidth]{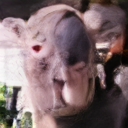}}
        \put(25, 75){\includegraphics[width=0.25\linewidth]{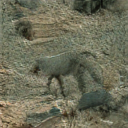}}
        \put(50, 75){\includegraphics[width=0.25\linewidth]{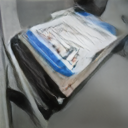}}
        \put(75, 75){\includegraphics[width=0.25\linewidth]{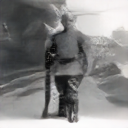}}
    
        \put(0, 50){\includegraphics[width=0.25\linewidth]{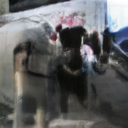}}
        \put(25, 50){\includegraphics[width=0.25\linewidth]{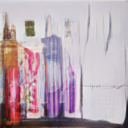}}
        \put(50, 50){\includegraphics[width=0.25\linewidth]{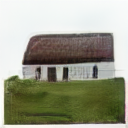}}
        \put(75, 50){\includegraphics[width=0.25\linewidth]{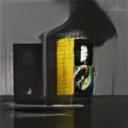}}
    
        \put(0, 25){\includegraphics[width=0.25\linewidth]{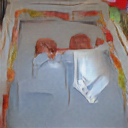}}
        \put(25, 25){\includegraphics[width=0.25\linewidth]{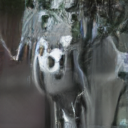}}
        \put(50, 25){\includegraphics[width=0.25\linewidth]{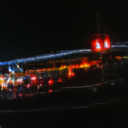}}
        \put(75, 25){\includegraphics[width=0.25\linewidth]{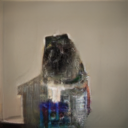}}
        
        \put(0, 0){\includegraphics[width=0.25\linewidth]{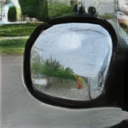}}
        \put(25, 0){\includegraphics[width=0.25\linewidth]{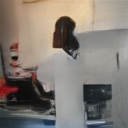}}
        \put(50, 0){\includegraphics[width=0.25\linewidth]{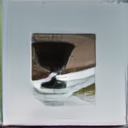}}
        \put(75, 0){\includegraphics[width=0.25\linewidth]{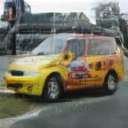}}
        \end{overpic}
        }

    \caption{
    Reconstruction results a whole batch of 16 when a $3$-bit QSGD quantizer $\pfunc_{\text{qsgd}}$ is applied.
    Compared with (a)~raw images, (b)~ROG reconstructed images are visually similar.}
    \label{fig:full_batch_qsgd}
\end{figure}

\begin{figure}[!tb]
   
    \begin{overpic}[width=\linewidth, height=\linewidth]{figs/4x4.pdf}
        \put(1, 95){\zoom{Batch}}
        \put(1, 92){\zoom{Size}}
        \put(2, 81){\zoom{16}}
        \put(2, 59){\zoom{32}}
        \put(2, 36){\zoom{64}}
        \put(2, 13){\zoom{128}}

        \put(15, 95){\zoom{Raw Image}}
        \put(35, 95){\zoom{Best Result}}
        \put(60, 95){\zoom{Raw Image}}
        \put(80, 95){\zoom{Worst Result}}
        
        \put(12, 71){\includegraphics[width=0.21\linewidth]{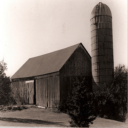}}
        \put(33.5, 71){\includegraphics[width=0.21\linewidth]{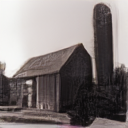}}
        \put(57.5, 71){\includegraphics[width=0.21\linewidth]{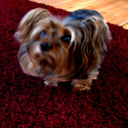}}
        \put(79, 71){\includegraphics[width=0.21\linewidth]{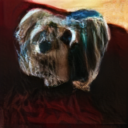}}
    
        \put(12, 49){\includegraphics[width=0.21\linewidth]{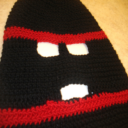}}
        \put(33.5, 49){\includegraphics[width=0.21\linewidth]{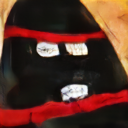}}
        \put(57.5, 49){\includegraphics[width=0.21\linewidth]{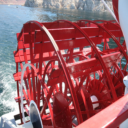}}
        \put(79, 49){\includegraphics[width=0.21\linewidth]{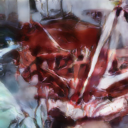}}
    
        \put(12, 27){\includegraphics[width=0.21\linewidth]{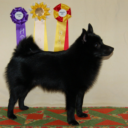}}
        \put(33.5, 27){\includegraphics[width=0.21\linewidth]{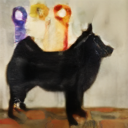}}
        \put(57.5, 27){\includegraphics[width=0.21\linewidth]{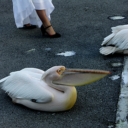}}
        \put(79, 27){\includegraphics[width=0.21\linewidth]{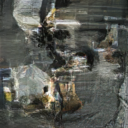}}
    
        \put(12, 5){\includegraphics[width=0.21\linewidth]{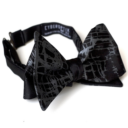}}
        \put(33.5, 5){\includegraphics[width=0.21\linewidth]{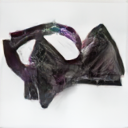}}
        \put(57.5, 5){\includegraphics[width=0.21\linewidth]{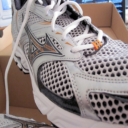}}
        \put(79, 5){\includegraphics[width=0.21\linewidth]{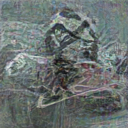}}
    \end{overpic}
    \vspace*{-30pt}
    \caption{
    Comparison between the best and the worst reconstructed images in a full batch. 
    The best reconstructed images convey meaningful pictorial information, whereas the worst reconstructed images are almost unrecognizable.
    \label{fig:best_worst}}
\end{figure}

\begin{figure}
    
\subcaptionbox{}[0.495\linewidth]{
    \begin{overpic}[width=\linewidth, height=\linewidth]{figs/4x4.pdf}
    \put(0, 75){\includegraphics[width=0.25\linewidth]{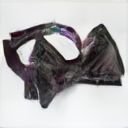}}
    \put(25, 75){\includegraphics[width=0.25\linewidth]{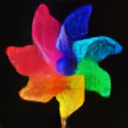}}
    \put(50, 75){\includegraphics[width=0.25\linewidth]{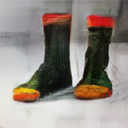}}
    \put(75, 75){\includegraphics[width=0.25\linewidth]{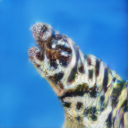}}

    \put(0, 50){\includegraphics[width=0.25\linewidth]{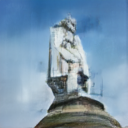}}
    \put(25, 50){\includegraphics[width=0.25\linewidth]{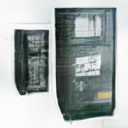}}
    \put(50, 50){\includegraphics[width=0.25\linewidth]{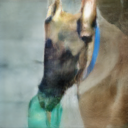}}
    \put(75, 50){\includegraphics[width=0.25\linewidth]{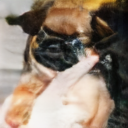}}

    \put(0, 25){\includegraphics[width=0.25\linewidth]{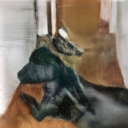}}
    \put(25, 25){\includegraphics[width=0.25\linewidth]{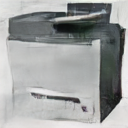}}
    \put(50, 25){\includegraphics[width=0.25\linewidth]{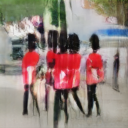}}
    \put(75, 25){\includegraphics[width=0.25\linewidth]{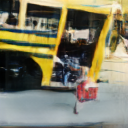}}
    
    \put(0, 0){\includegraphics[width=0.25\linewidth]{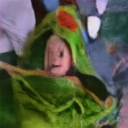}}
    \put(25, 0){\includegraphics[width=0.25\linewidth]{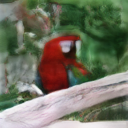}}
    \put(50, 0){\includegraphics[width=0.25\linewidth]{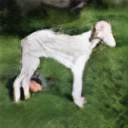}}
    \put(75, 0){\includegraphics[width=0.25\linewidth]{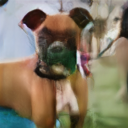}}
    \end{overpic}
}
\subcaptionbox{}[0.495\linewidth]{
    \begin{overpic}[width=\linewidth, height=\linewidth]{figs/4x4.pdf}
    \put(0, 75){\includegraphics[width=0.25\linewidth]{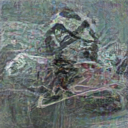}}
    \put(25, 75){\includegraphics[width=0.25\linewidth]{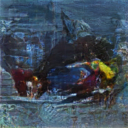}}
    \put(50, 75){\includegraphics[width=0.25\linewidth]{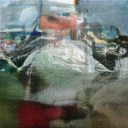}}
    \put(75, 75){\includegraphics[width=0.25\linewidth]{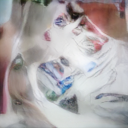}}

    \put(0, 50){\includegraphics[width=0.25\linewidth]{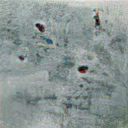}}
    \put(25, 50){\includegraphics[width=0.25\linewidth]{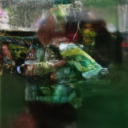}}
    \put(50, 50){\includegraphics[width=0.25\linewidth]{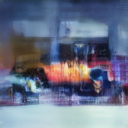}}
    \put(75, 50){\includegraphics[width=0.25\linewidth]{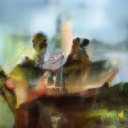}}

    \put(0, 25){\includegraphics[width=0.25\linewidth]{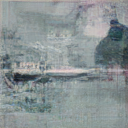}}
    \put(25, 25){\includegraphics[width=0.25\linewidth]{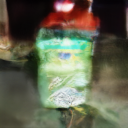}}
    \put(50, 25){\includegraphics[width=0.25\linewidth]{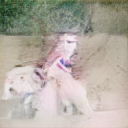}}
    \put(75, 25){\includegraphics[width=0.25\linewidth]{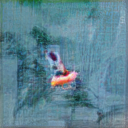}}
    
    \put(0, 0){\includegraphics[width=0.25\linewidth]{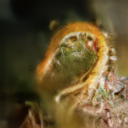}}
    \put(25, 0){\includegraphics[width=0.25\linewidth]{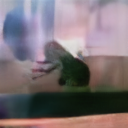}}
    \put(50, 0){\includegraphics[width=0.25\linewidth]{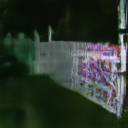}}
    \put(75, 0){\includegraphics[width=0.25\linewidth]{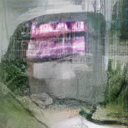}}
    \end{overpic}
}
\caption{
    (a)~The best 16 reconstructed images and (b) the worst 16 reconstructed images from a full batch of size 128.
\label{fig:best_worst_128}}
\end{figure}

We provide more reconstruction results on the whole batch in addition to those given in Section~\ref{section:rog} and Figure~\ref{fig:full_batch}. 
The reconstruction of a batch size of $16$ under a $3$-bit QSGD quantizer is shown in \paperfig{fig:full_batch_qsgd}. 
We compare the best and worst reconstruction results in \paperfig{fig:best_worst} when different batch sizes are used.
The best and worst results are selected based on LPIPS values. 
In \paperfig{fig:best_worst_128}, we show the best $16$ and worst $16$ reconstructed images when the batch size is set to $128$. 
The best reconstructed images convey meaningful pictorial information, whereas the worst reconstructed images are almost unrecognizable.

\section{Additional Experiments}\label{section:add_results}

\begin{figure}[!tb]
    \subcaptionbox{\label{subfig:fullbatch_invgrad}}[0.495\linewidth]{
        \begin{overpic}[width=\linewidth, height=\linewidth]{figs/4x4.pdf}
        \put(0, 75){\includegraphics[width=0.25\linewidth]{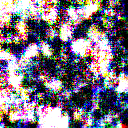}}
        \put(25, 75){\includegraphics[width=0.25\linewidth]{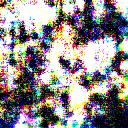}}
        \put(50, 75){\includegraphics[width=0.25\linewidth]{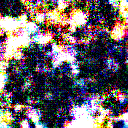}}
        \put(75, 75){\includegraphics[width=0.25\linewidth]{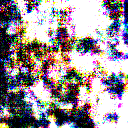}}

        \put(0, 50){\includegraphics[width=0.25\linewidth]{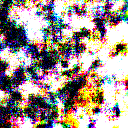}}
        \put(25, 50){\includegraphics[width=0.25\linewidth]{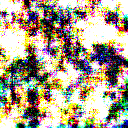}}
        \put(50, 50){\includegraphics[width=0.25\linewidth]{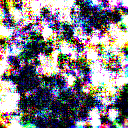}}
        \put(75, 50){\includegraphics[width=0.25\linewidth]{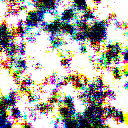}}

        \put(0, 25){\includegraphics[width=0.25\linewidth]{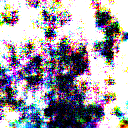}}
        \put(25, 25){\includegraphics[width=0.25\linewidth]{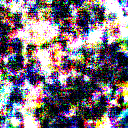}}
        \put(50, 25){\includegraphics[width=0.25\linewidth]{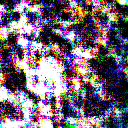}}
        \put(75, 25){\includegraphics[width=0.25\linewidth]{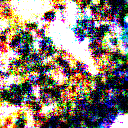}}
        
        \put(0, 0){\includegraphics[width=0.25\linewidth]{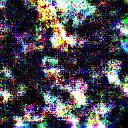}}
        \put(25, 0){\includegraphics[width=0.25\linewidth]{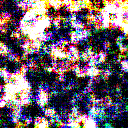}}
        \put(50, 0){\includegraphics[width=0.25\linewidth]{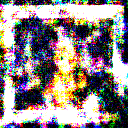}}
        \put(75, 0){\includegraphics[width=0.25\linewidth]{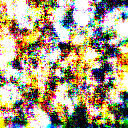}}
        \end{overpic}
    }
    \subcaptionbox{\label{subfig:fullbatch_dlg_post}}[0.495\linewidth]{
        \begin{overpic}[width=\linewidth, height=\linewidth]{figs/4x4.pdf}
        \put(0, 75){\includegraphics[width=0.25\linewidth]{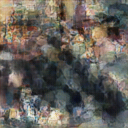}}
        \put(25, 75){\includegraphics[width=0.25\linewidth]{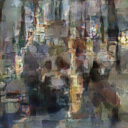}}
        \put(50, 75){\includegraphics[width=0.25\linewidth]{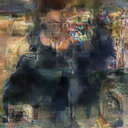}}
        \put(75, 75){\includegraphics[width=0.25\linewidth]{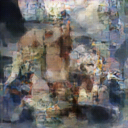}}
    
        \put(0, 50){\includegraphics[width=0.25\linewidth]{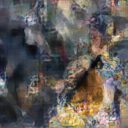}}
        \put(25, 50){\includegraphics[width=0.25\linewidth]{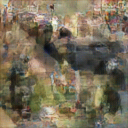}}
        \put(50, 50){\includegraphics[width=0.25\linewidth]{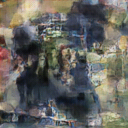}}
        \put(75, 50){\includegraphics[width=0.25\linewidth]{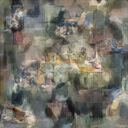}}
    
        \put(0, 25){\includegraphics[width=0.25\linewidth]{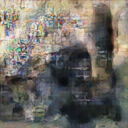}}
        \put(25, 25){\includegraphics[width=0.25\linewidth]{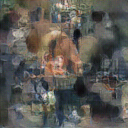}}
        \put(50, 25){\includegraphics[width=0.25\linewidth]{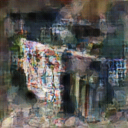}}
        \put(75, 25){\includegraphics[width=0.25\linewidth]{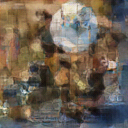}}
        
        \put(0, 0){\includegraphics[width=0.25\linewidth]{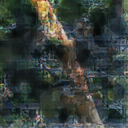}}
        \put(25, 0){\includegraphics[width=0.25\linewidth]{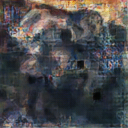}}
        \put(50, 0){\includegraphics[width=0.25\linewidth]{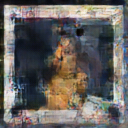}}
        \put(75, 0){\includegraphics[width=0.25\linewidth]{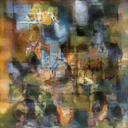}}
        \end{overpic}
    }

    \caption{
    Reconstruction results on a whole batch of 16 with the attack methods (a)~InvertGrad with a postprocessing module and (b)~DLG with a postprocessing module. 
    \label{fig:full_batch_baseline}}
\end{figure}

We discuss some additional simulation results in this section. 
First, we demonstrate the full batch reconstruction with InvertGrad and DLG in Figure~\ref{fig:full_batch_baseline}. 
For InvertGrad, we use the Adam optimizer and set the number of iterations to $24$k. 
For DLG, we use L-BFGS optimizer and set the number of iterations to $300$. 
We use the original implementations provided by~\cite{zhu2019deep, geiping2020inverting}. 

\begin{figure}[!tb]
    \begin{overpic}[width=\linewidth]{figs/4x4.pdf}
    \put(17, 69){\includegraphics[width=0.2\linewidth]{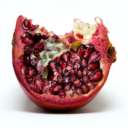}}
    \put(38, 69){\includegraphics[width=0.2\linewidth]{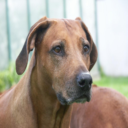}}
    \put(59, 69){\includegraphics[width=0.2\linewidth]{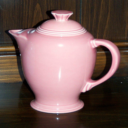}}
    \put(80, 69){\includegraphics[width=0.2\linewidth]{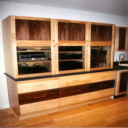}}

    \put(17, 67){\includegraphics[width=0.83\linewidth, height=1.3pt]{figs/lines/line.png}}

    \put(17, 45){\includegraphics[width=0.2\linewidth]{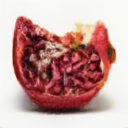}}
    \put(38, 45){\includegraphics[width=0.2\linewidth]{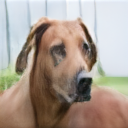}}
    \put(59, 45){\includegraphics[width=0.2\linewidth]{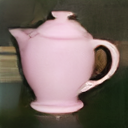}}
    \put(80, 45){\includegraphics[width=0.2\linewidth]{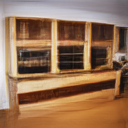}}

    \put(17, 24){\includegraphics[width=0.2\linewidth]{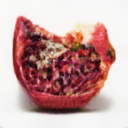}}
    \put(38, 24){\includegraphics[width=0.2\linewidth]{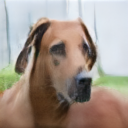}}
    \put(59, 24){\includegraphics[width=0.2\linewidth]{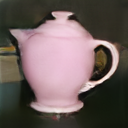}}
    \put(80, 24){\includegraphics[width=0.2\linewidth]{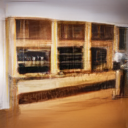}}

    \put(17, 3){\includegraphics[width=0.2\linewidth]{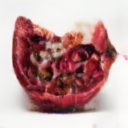}}
    \put(38, 3){\includegraphics[width=0.2\linewidth]{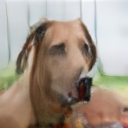}}
    \put(59, 3){\includegraphics[width=0.2\linewidth]{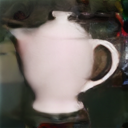}}
    \put(80, 3){\includegraphics[width=0.2\linewidth]{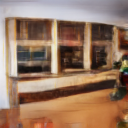}}

    \put(0, 76){\zoom{Raw Image}}
    \put(0, 55){\zoom{LeNet}}
    \put(0, 51){\zoom{LPIPS $0.173$}}
    \put(0, 35.5){\zoom{VGG-7}}
    \put(0, 31.5){\zoom{LPIPS $0.193$}}
    \put(0, 16){\zoom{ResNet-18}}
    \put(0, 12){\zoom{LPIPS $0.291$}}
    \end{overpic}
    \vspace*{-20pt}
    \caption{
    Attack against different neural network architectures. 
    All reconstructed images are visually similar to raw images. 
    \label{fig:attack_arch}}
\end{figure}

\highlight{Neural Network Architectures }
In this experiment, we study the impact of different neural network architectures on the attack scheme. 
We choose the LeNet, VGG-7, and ResNet-18 and demonstrate the reconstruction results in Figure~\ref{fig:attack_arch}. 
LeNet is a five-layer convolutional network with $5\times5$ kernels, and VGG-7 is a seven-layer convolutional network with $3\times3$ kernels.
ResNet-18 is an eighteen-layer residual network with skip connections. 
Intuitively, the increased number of nodes in a neural network appears to give the adversary more advantages, as the number of known conditions will also increase.  
On the other hand, more sophisticated neural network architecture will have an increased nonlinearity, thus impeding a successful reconstruction.  
In Figure~\ref{fig:attack_arch}, we can observe that all reconstructed images under different neural network architectures are visually similar. 
Our observation is consistent with the prior study that changing the neural network architecture may not directly affect the privacy protection level. 
A more extensive study concerning neural network architectures is warranted in future work.

\end{document}